\documentclass[apj,twocolumn]{emulateapj_mod}
\usepackage{epsfig,apjfonts,mathptmx}

\newcommand{\bsex}{\hbox{\tt SExtractor}}
\newcommand{\etal}{\hbox{et al.\,}}
\newcommand{\Kv}  {\hbox{$K_{\rm Vega}$}}
\newcommand{\sbzks}{\hbox{sBzKs}}
\newcommand{\pegs} {\hbox{pBzKs}}
\newcommand{\gsim}{\lower.5ex\hbox{$\; \buildrel > \over \sim \;$}}
\newcommand{\lsim}{\lower.5ex\hbox{$\; \buildrel < \over \sim \;$}}
\newcommand{\psim}{\lower.5ex\hbox{$\; \buildrel \propto \over \sim \;$}}
\def\gtsima{$\; \buildrel > \over \sim \;$}
\def\ltsima{$\; \buildrel < \over \sim \;$}
\def\prosima{$\; \buildrel \propto \over \sim \;$}
\def\gsim{\lower.5ex\hbox{\gtsima}}
\def\lsim{\lower.5ex\hbox{\ltsima}}
\def\simgt{\lower.5ex\hbox{\gtsima}}
\def\simlt{\lower.5ex\hbox{\ltsima}}
\def\simpr{\lower.5ex\hbox{\prosima}}
\def\h1{$h^{-1}$}
\def\eeq{\end{equation}}
\def\beq{\begin{equation}}

\shorttitle{Survey for High-Redshift Massive Galaxies I.}
\shortauthors{X. Kong et al.}

\begin{document}

\title{A wide area survey for high-redshift massive galaxies. 
I. Number counts and clustering of
B$\lowercase{\rm z}$K$\lowercase{\rm s}$
and ERO$\lowercase{\rm s}^1$}

\author{
X. Kong\altaffilmark{1,4},   
E. Daddi\altaffilmark{2},   
N. Arimoto\altaffilmark{1},
A. Renzini\altaffilmark{3,5},
T. Broadhurst\altaffilmark{6},
A. Cimatti\altaffilmark{7},
C. Ikuta\altaffilmark{1},   
K. Ohta\altaffilmark{8},      
L. da Costa\altaffilmark{5}, 
L.~F. Olsen\altaffilmark{9},
M. Onodera\altaffilmark{1,10},
N. Tamura\altaffilmark{11}
}

\altaffiltext{1}{Based on data collected at Subaru Telescope
(program S02B-101), which is operated by the National 
Astronomical Observatory of Japan. Also based on data 
collected at the New Technology Telescope (program IDs 
169.A-0725; 67.A-0244(A); 164.O-0561), which is operated by 
the European Southern Observatory.}  

\affil{$^1$Optical and Infrared Astronomy Division, National 
Astronomical Observatory, 
Mitaka, 
Tokyo 181-8588, Japan}
\affil{$^2${\em Spitzer Fellow}, National Optical Astronomy
Observatory, P.O. Box 26732, Tucson, AZ 85726, USA}
\affil{$^3$Osservatorio Astronomico di Padova, Vicolo 
dell'Osservatorio 5, I-35122 Padova, Italy}
\affil{$^4$Center for Astrophysics, University of Science and 
Technology of China, Hefei 230026, China}
\affil{$^5$European Southern Observatory, Karl-Schwarzschild-Str.
 2, D-85748 Garching, Germany} 
\affil{$^6$School of Physics and Astronomy, Tel Aviv University, 
Tel Aviv 69978, Israel} 
\affil{$^7$INAF--Osservatorio Astrofisico di Arcetri, Largo E. 
Fermi 5, I-50125 Firenze, Italy}
\affil{$^8$Department of Astronomy, Kyoto University, Kyoto 
606-8502, Japan}
\affil{$^9$Observatoire de la C\^ote d'Azur, Laboratoire 
Cassiop\'ee, BP 4229, 06304 Nice Cedex 4, France}
\affil{$^{10}$Department of Astronomy, School of Science, University 
of Tokyo, Japan}
\affil{$^{11}$Department of Physics, University of Durham, South 
Road, Durham, DH1 3LE, UK}

\begin{abstract}
We present the results of a deep, wide-area, optical and near-IR
survey of massive high-redshift galaxies.
The Prime Focus Camera (Suprime-Cam) on the Subaru telescope was 
used to obtain $BRIz'$ imaging over $2 \times 940$ arcmin$^2$
fields, while $JK_s$ imaging was provided by the SOFI camera at 
the New Technology Telescope (NTT) for a subset of the area, partly 
from the ESO Imaging Survey (EIS). 
In this paper, we report on the properties of $K$-band--selected 
galaxies, identified from a total area of $\sim 920$ arcmin$^2$ to 
$\Kv=19$, of which 320 arcmin$^2$ are complete to $\Kv=20$.  
The $BzK$ selection technique was used to assemble complete samples 
of about 500 candidate massive star-forming galaxies (\sbzks) and 
about 160 candidate massive passively evolving galaxies (\pegs) at 
$1.4 \lsim z \lsim 2.5$; and the $(R-K)_{\rm Vega} > 5$ color criterion 
was used to assemble a sample of about 850 extremely red objects 
(EROs).
We accurately measure surface densities of $1.20\pm0.05$ arcmin$^{-2}$ 
and $0.38\pm0.03$ arcmin$^{-2}$ for the \sbzks\ and the \pegs,
respectively.
Both \sbzks\ and \pegs\ are strongly clustered, at a level at least 
comparable to that of EROs, with \pegs\ appearing more clustered 
than \sbzks.
We estimate the reddening, star formation rates (SFRs) and stellar 
masses ($M_*$) for the ensemble of \sbzks, confirming that to 
$\Kv \sim 20$ typical (median) values are $M_*\sim10^{11}\ M_\odot$, 
$SFR \sim 190\ M_{\odot} yr^{-1}$, and $E(B-V)\sim0.44$. 
A correlation is detected such that the most massive galaxies at 
$z\sim2$ are also the most actively star-forming, an effect that 
can be seen as a manifestation of {\it downsizing} at early epochs.
The space density of massive \pegs\ at $z\sim 1.4-2$ that we 
derive is 20\%$\pm7$\% that of similarly massive early-type galaxies 
at $z\sim 0$.
Adding this space density to that of our massive star forming class, 
\sbzks, in the same redshift range produces a closer comparison with 
the local early-type galaxy population, naturally implying that we 
are detecting star formation in a sizable fraction of massive 
galaxies at $z > 1.4$, which has been quenched by the present day.
Follow-up optical and near infrared spectroscopy is in progress at 
the ESO Very Large Telescope (VLT) and at the Subaru telescope, in 
order to elucidate more thoroughly the formation and evolution of 
massive galaxies.
\end{abstract}


\section{Introduction}\label{sec:int}

Despite the recent extraordinary progress in observational cosmology
and the successful convergence on a single cosmological model, galaxy
formation and evolution largely remain an open issue. One
critical aspect is how and when the present-day most massive galaxies
(e.g. elliptical galaxies and bulges with $M_*\gsim 10^{11}M_\odot$)
were built up and what type of evolution characterized their growth
over cosmic time (e.g., Cimatti \etal 2004; Glazebrook \etal 2004,
and references therein). 
Indeed, various current renditions of the $\Lambda$CDM hierarchical
merging paradigm differ enormously in this respect, with some models
predicting the virtually complete disappearance of such galaxies by
$z=1-2$ (e.g., Cole \etal 2000; Menci \etal 2002; Somerville 2004a)
and other models predicting a quite mild evolution, more in line with
observations (e.g., Nagamine \etal 2001; 2005; Granato \etal 2004; 
Somerville \etal 2004b; a direct comparison of such models can be 
found in Fig. 9 of Fontana \etal 2004). Moreover, models that 
provide an acceptable fit to the galaxy stellar mass function at 
$z>1$ may differ considerably in the actual properties of the 
galaxies with $M_*\gsim 10^{11}M_\odot$ at $z\gsim 1$, with some 
models predicting very few, if any, passively evolving galaxies at 
these redshifts, at variance with recent findings (Cimatti \etal 
2004; McCarthy \etal 2004; Daddi \etal 2005a; Saracco \etal 2005).

While various $\Lambda$CDM models may agree with each other at 
$z\sim 0$ (where they all are tuned) their dramatic divergence with
increasing redshift gives us powerful leverage to restrict the
choice among them, thus aiding understanding of the physics of 
galaxy formation and evolution. Hence, a direct observational mapping 
of galaxy evolution through cosmic time is particularly important and 
rewarding, especially if a significant number of massive galaxies at 
$1\lsim z\lsim 3$ can be identified and studied. 
In this regard, the critical questions concern the evolution with 
redshift of the number density of massive galaxies and their star 
formation histories, as reflected by their colors and spectral energy 
distributions (SEDs). These questions have just started to be addressed by 
various spectroscopy projects, such as the K20 survey (Cimatti \etal 
2002a, 52 arcmin$^2$), the Hubble Deep Fields (HDFs; Ferguson \etal 2000, 
5.3 arcmin$^2$ in the HDF-North and 4.4 arcmin$^2$ in the HDF-South), the Great 
Observatories Origins Deep Survey (GOODS; Giavalisco \etal 2004, 320 
arcmin$^2$ in the North and South fields combined) the HST/ACS Ultra 
Deep Field (S. Beckwith \etal 2006, in preparation; 12 arcmin$^2$), the 
Gemini Deep Deep Survey (Abraham \etal 2004, 121 arcmin$^2$), and 
the extension down to $z\sim 2$ of the Lyman break galaxy (LBG) project 
(Steidel \etal 2004, $\sim 100$ arcmin$^2$).
However, massive galaxies are quite rare and likely highly clustered 
at all redshifts, and hence small areas such as those explored so far are 
subject to large cosmic variance (Daddi \etal 2000; Bell \etal 2004;
Somerville \etal 2004c). 
Therefore, although these observation have demonstrated that old, 
passive and massive galaxies do exist in the field out to $z \sim 2$, 
it remains to be firmly established how their number and evolutionary 
properties evolve with redshift up to $z\sim 2$ and beyond.

To make a major step forward we are undertaking fairly deep,
wide-field imaging with the Suprime-Cam on Subaru of two fields of 
940 arcmin$^2$ each for part of which near-IR data are available from  
ESO New Technology Telescope (NTT) observations. 
The extensive imaging has supported the spectroscopic follow-up with 
the VLT and the Subaru telescopes, for which part of the data have 
already been secured. The prime aim of this survey is to understand 
how and when the present-day massive galaxies formed, and to this end, 
the imaging observations have been optimized for the use of
optical/near-IR multi-color selection criteria to identify both
star-forming and passive galaxies at $z\approx 2$.

Color criteria are quite efficient in singling out high redshift 
galaxies. The best-known example is the dropout technique for 
selecting LBGs (Steidel \etal 1996). Besides 
targeting LBGs, color criteria have also been used to search for 
passively evolving galaxies at high redshifts, such as extremely 
red objects (EROs) at redshifts $z\sim 1$ (Thompson \etal 1999; 
McCarthy 2004) and distant red galaxies (DRGs) at redshifts 
$z\gsim2$ (Franx \etal 2003).

Recently, using the highly complete spectroscopic redshift database 
of the K20 survey, Daddi \etal (2004a) introduced a new 
criterion for obtaining virtually complete samples of galaxies
in the redshift range $1.4\lsim z \lsim 2.5$, based on $B$, $z$ and 
$K_s$\footnote{hereafter $K$ band for short}
imaging: 
star-forming galaxies are identified requiring
$BzK=(z-K)_{\rm AB}-(B-z)_{\rm AB}>-0.2$
(for convenience, we use the term \sbzks\ for galaxies selected in 
this way); and passively evolving galaxies at $z\gsim1.4$ requiring 
$BzK<-0.2$ and $(z-K)_{\rm AB}>2.5$ (hereafter \pegs).
This criterion is reddening independent for star-forming galaxies in 
the selected redshift range, thus allowing us also to select the reddest 
most dust-extinguished galaxies, together with those that are old 
and passively evolving. 
This should allow for a relatively unbiased selection of $z\sim2$ 
galaxies within the magnitude limit of the samples studied.

In this paper observations, data reduction and galaxy photometry are 
described, together with the first results on K-band selected samples 
of distant, high redshift massive galaxies.
Compared to optical, the near-IR selection (in particular in the $K$ 
band) offers several advantages, including the relative insensitivity 
of the k-corrections to galaxy type, even at high redshift, the 
less severe dust extinction effects, the weaker dependence on the 
instantaneous star formation activity, and a tighter correlation with 
the stellar mass of the galaxies. 
Therefore, the study of faint galaxy samples selected in the 
near-infrared have long been recognized as ideal tools to study the 
process of mass assembly at high redshift (Broadhurst \etal 1992; 
Kauffmann \& Charlot 1998; Cimatti \etal 2002a). 

The paper is organized as follows:
Section 2 describes the observations and the data reduction. 
Section 3 discusses the photometric calibration of the images. 
Section 4 presents the selection and number counts for EROs, 
\sbzks, and \pegs.
Section 5 presents the analysis of the clustering of field galaxies, 
EROs, \sbzks, and \pegs.
The properties of \sbzks\ are presented in Section 6. 
Finally, a brief summary is presented in Section 7. 
Throughout the paper, we use  the Salpeter IMF extending between 0.1 
and 100 $M_\odot$ and a cosmology with $\Omega_\Lambda =0.7, 
\Omega_M = 0.3$, and $h = H_0$(km s$^{-1}$ Mpc$^{-1}$)$/100=0.71$. 
For the sake of comparison with previous works, magnitudes and colors
 in both AB and Vega systems have to be used.\footnote{ The relevant 
conversions between Vega and AB magnitudes for this 
paper are $B_{\rm AB}=B_{\rm Vega}-0.08$, 
$R_{\rm AB}=R_{\rm Vega}+0.22$, $z_{\rm AB}=z_{\rm Vega}+0.53$, and
$K_{\rm AB}=\Kv+1.87$.}

\section{Observations}\label{sec:obs}

\begin{figure*} 
\centering
\includegraphics[width=0.45\textwidth]{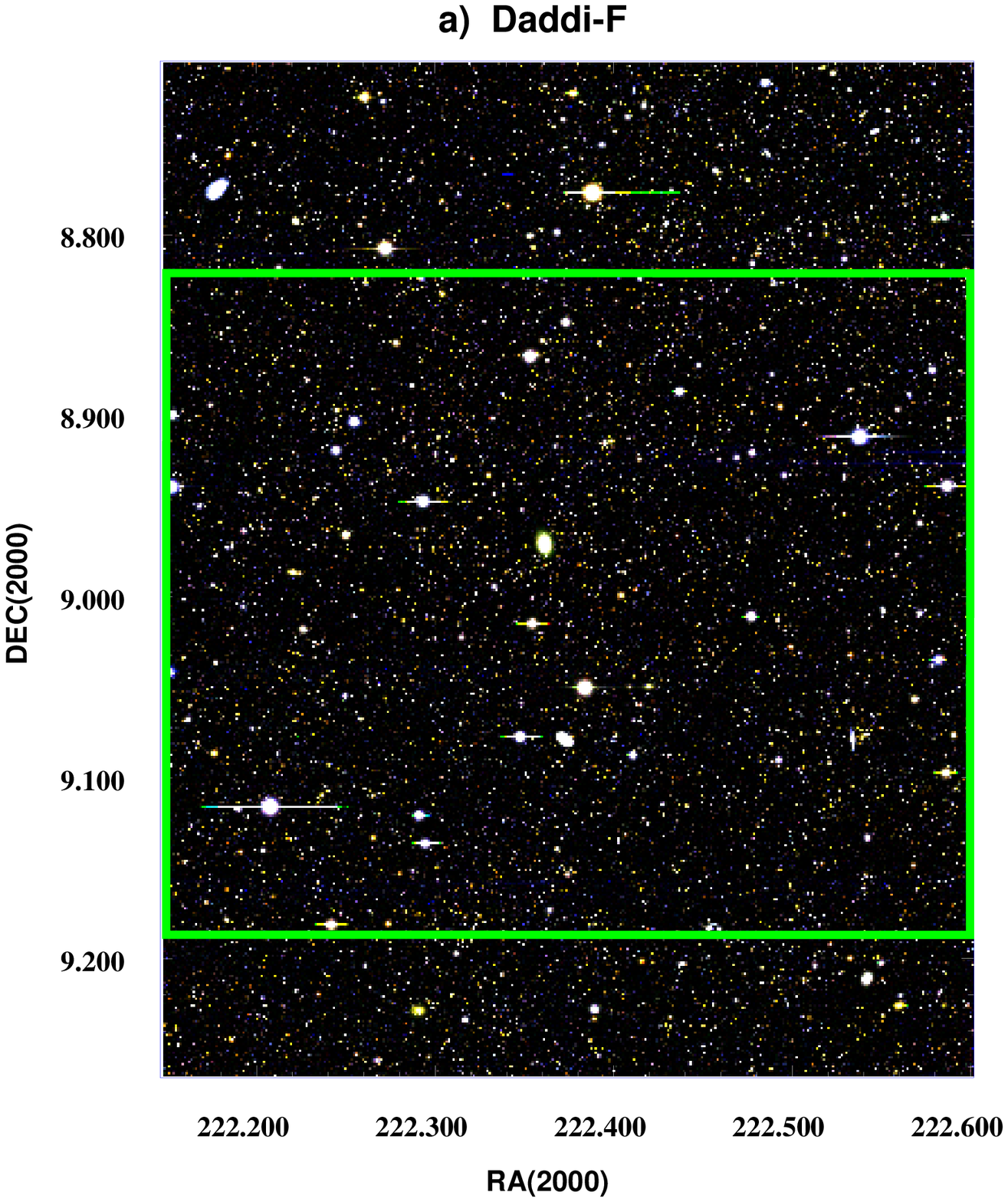} 
\includegraphics[width=0.54\textwidth]{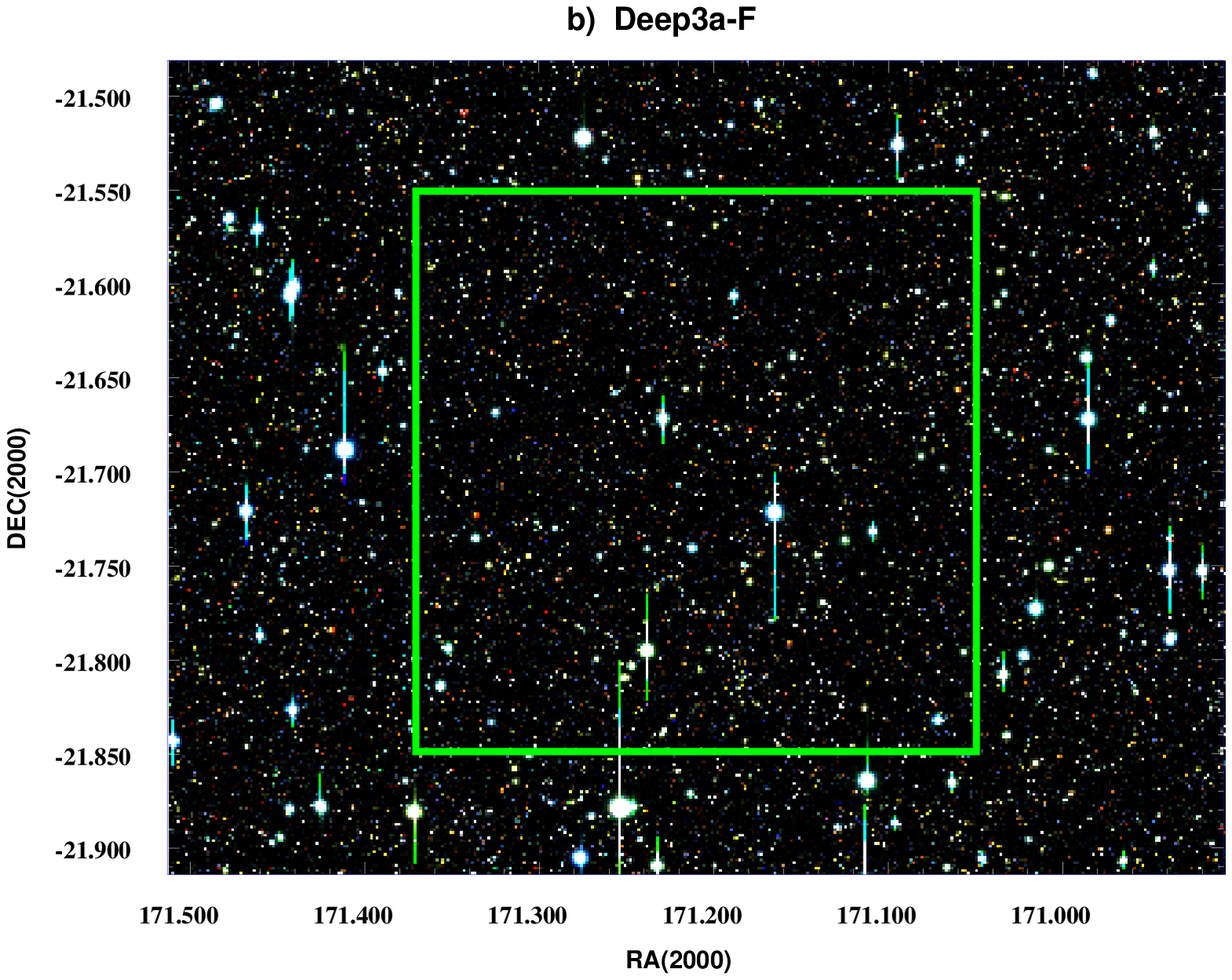} 
\caption{
Composite pseudo-color images of Daddi-F (a) and Deep3a-F (b). 
The RGB colors are assigned to $z$-, $I$-, and $B$-band images,
940 arcmin$^2$, respectively. 
The green area outlined near the center of the images is the field where 
$K$-band images have been obtained by NTT (600 arcmin$^2$ area for 
Daddi-F and 320 arcmin$^2$ area for Deep3a-F).  
}
\label{fig:area}
\end{figure*}

Two widely separated fields were imaged as a part of our survey: 
one centered at  
$\alpha$(J2000)$ = 11^h24^m50^s$, 
$\delta$(J2000)$=-21^{\circ}42^{\prime}00^{\prime \prime}$ 
(hereafter Deep3a-F), and the the second, the so-called 
``Daddi field'' (hereafter Daddi-F; Daddi \etal 2000) 
centered at $\alpha$(J2000)$ = 14^h49^m29^s$, 
$\delta$(J2000)$ = 09^{\circ}00^{\prime}00^{\prime \prime}$. Details 
of the optical and near-IR observations are shown in Table 1.
Figure~\ref{fig:area} shows the layout of the two areas observed.

\begin{table*}
\centering
\caption{Journal of observations.}
\label{tab:obs}
\begin{tabular*}{0.9\textwidth}{@{\extracolsep{\fill}}llrcccr}
\tableline
\tableline
Filter  &Telescope &Obs. date   &Exps.\tablenotemark{a} &
Seeing&$m_{lim}$\tablenotemark{b} &Area\\
        &          &        &(sec) &($''$)&(mag) &(arcmin$^2$)\\
\tableline
\\
\multicolumn{7}{c}{Deep3a-F}\\	
\tableline
{\it B}    &Subaru &Mar.5,   03      &3900  &0.77 &27.4  &940       \\
{\it R$_c$}    &Subaru &Mar.4-5, 03      &7320  &0.85 &26.9  &940       \\
{\it I}    &Subaru &Mar.4-5, 03      &5700  &0.77 &26.5  &940       \\
{\it z$'$} &Subaru &Mar.4-5, 03      &9900  &0.80 &26.0  &940       \\
{\it J}    &NTT    &Jan.00-Feb.01    &3600  &0.76 &23.4  &320       \\
{\it $K_s$}&NTT    &Jan.00-Feb.01    &4800  &0.76 &22.7  &320       \\
\tableline
\\
\multicolumn{7}{c}{Daddi-F}\\	
\tableline
{\it B}    &Subaru &Mar. 5, 03       &1500  &0.75 &27.0  &940       \\
{\it R}\tablenotemark{c}&WHT &May 19-21, 98 &3600 &0.70 &25.6  &715 \\
{\it I}    &Subaru &Mar. 5, 03       &1800  &0.90 &26.0  &940       \\
{\it z$'$} &Subaru &Mar. 4-5, 03     &2610  &0.80 &25.5  &940       \\
{\it $K_s$}&NTT    &Mar. 27-30, 99   &720   &0.90 &21.5  &600       \\
\tableline
\end{tabular*}
\tablenotetext{a}{Exposure value for K-band images are ``typical 
values"; see text.}
\tablenotetext{b}{The limiting magnitude (in AB) is defined as 
the brightness corresponding to 5 $\sigma$ on a 2$''$ diameter 
aperture.}
\tablenotetext{c}{$R-$ and $K$-band data of Daddi-F are described in 
Daddi \etal (2000).}
\end{table*}

\subsection{Near-IR imaging and data reduction}

Infrared observations in the near-infrared passband $J$ and $K_s$ 
were obtained using the SOFI camera 
(Moorwood, Cuby \& Lidman 1998) mounted on the New Technology 
Telescope (NTT) at La~Silla. SOFI is equipped with a Rockwell 
1024$^2$ detector, which, when used together with its large field 
objective, provides images with a pixel scale of $0''.29$ and a 
field of view of $\sim 4.9\times 4.9$ arcmin$^2$. 

Deep3a-F is part of the ESO Deep Public Survey (DPS) carried out by 
the ESO Imaging Survey (EIS) program  (Renzini \& da Costa 1999) 
(see http://www.eso.org/science/eis/). 
The Deep3a SOFI observations cover a total area of about 920 
arcmin$^2$ in the K band, most at the relatively shallow limits of
 $\Kv\sim19.0$--19.5. About 320 arcmin$^2$, the region used in 
the present paper, have much deeper integrations with a minimum 
3600s per sky pixel (and up to 2 hr) reaching to $\Kv\gsim20$ 
 and $J_{\rm Vega}\gsim22$.
The NTT $J$- and $K$-band images of Deep3a-F were retrieved from the 
ESO Science Archive and reduced using the EIS/MVM pipeline for 
automated reduction of optical/infrared images (Vandame 2002). 
The software produces fully reduced images and weight-maps carrying 
out bias subtraction, flat-fielding, de-fringing, background 
subtraction, first-order pixel-based image stacking (allowing for 
translation, rotation and stretching of the image) and astrometric 
calibration. Mosaicking of individual SOFI fields was based on the 
astrometric solution.

Photometric calibration was performed using standard stars from 
Persson \etal (1998), and the calibration performed as linear fits 
in airmass and color index whenever the airmass and color coverage 
allowed for it.
The reduced NTT $K$-band data and WHT $R$-band data for Daddi-F 
were taken from Daddi \etal (2000). The average seeing and the size 
of the final coadded images are reported in Table 1.

\subsection{Optical imaging and data reduction}

Deep optical imaging was obtained with the Prime Focus Camera on the 
Subaru Telescope, Suprime-Cam, which with its 10 $2$k $\times 4$k 
MIT/LL CCDs covers a contiguous area of $34'\times27'$ with a pixel 
scale of $0.''202$ pixel$^{-1}$ (Miyazaki \etal 2002). 
Deep3a-F was observed with the four standard broad-band filters, $B$, 
$R_c$ (hereafter $R$ band), $I$, and $z'$ (hereafter $z$ band) on the 
two nights of 2003 March 
4--5  with $0''.7 - 0''.9$ seeing. During the same nights also 
Daddi-F was imaged in $B$, $I$, and $z$ to a somewhat shallower 
magnitude limit to match the shallower $R$ and $K$ data from Daddi 
\etal (2000). A relatively long unit exposure time of several hundred 
seconds was used in order to reach background-noise-dominated levels. 
For this reason bright stars, which are saturated in the optical 
images, have been excluded by subsequent analysis.

During the same nights the photometric standard-star field SA95 
(Landolt 1992) was observed for $B-$, $-R$, and $I-$band flux 
calibration, and the SDSS standard-star fields SA95-190 and SA95-193 
were observed for $z$-band flux calibration (Smith \etal 2002).  

The Subaru imaging was reduced using the pipeline package SDFRED 
(Yagi \etal 2002; Ouchi \etal 2004).
The package includes overscan correction, bias subtraction, 
flat-fielding, correction for image distortion, PSF matching 
(by Gaussian smoothing), sky subtraction, and mosaicking.
Bias subtraction and flat fielding were processed in the same 
manner as for the conventional single chip CCD. 

In mosaicking, the relative positions (shifts and rotations) and 
relative throughput between frames taken with different CCDs and 
exposures are calculated using stars common to adjacent frames and 
running \bsex\ (Bertin \& Arnout 1996) with an S/N = 10 threshold.

\section{Photometry}\label{sec:photo}

\begin{figure*}
\centering 
\includegraphics[angle=-90,width=0.9\textwidth]{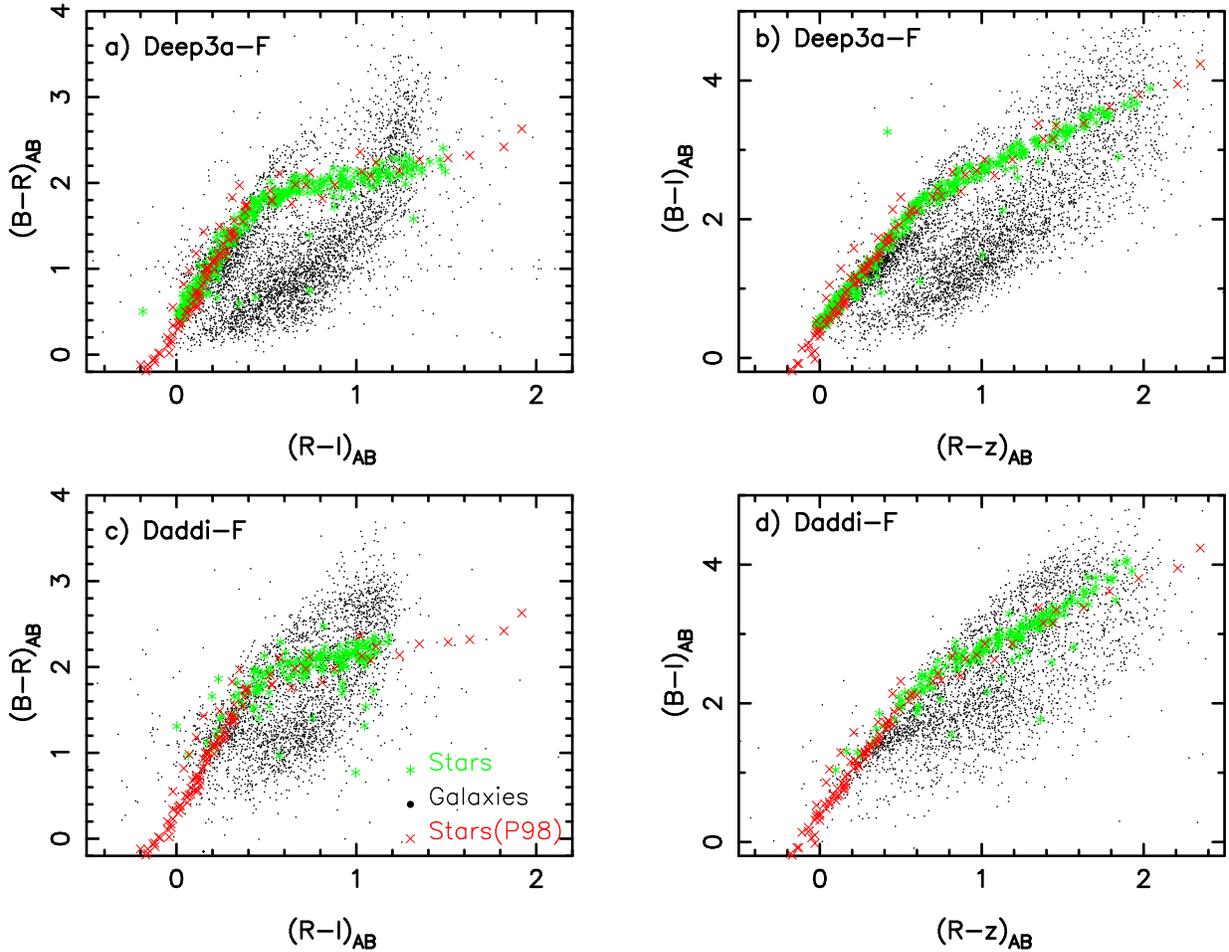}
\caption{
Optical two-color plots of $K$-selected objects in our survey:
the left panels show $B-R$ vs. $R-I$ colors; the right panels have
$B-I$ vs. $R-z$ colors (all in AB scale).
Galaxies are shown as filled points and stars with green asterisks
(based on the $BzK$ color star-galaxy seperation, Sect.~\ref{sec:S/G}).
[{\it See the electronic edition of the Journal for the color 
version of this figure.}]
}
\label{fig:ccd}
\end{figure*}

We obtained K-selected catalogs of objects in our survey by detecting 
sources in the K-band mosaics. For the Daddi-F we used the sample of
$K$-selected galaxies defined in Daddi \etal (2000).
\bsex\, (Bertin \& Arnouts 
1996) was used to perform the image analysis 
and source detection in Deep3a-F.
The total magnitudes were then defined as the brightest between the 
Kron automatic aperture magnitudes and the corrected aperture 
magnitude. 
Multicolor photometry in all the available bands was obtained by 
running \bsex\ in double image mode, after aligning all imaging to 
the K-band mosaic.
Colors were measured using 2$''$ diameter aperture magnitudes, corrected for 
the flux loss of stars.
The aperture corrections were estimated from the difference between 
the \bsex\, Kron automatic aperture magnitudes (MAG\_AUTO) and the 
2$\arcsec$ aperture magnitudes, resulting in a range of 
0.10 -- 0.30 mag, depending on the seeing.
All magnitudes were corrected for Galactic extinction ($A_B = 0.18$ 
and $0.13$ for Deep3a-F and Daddi-F, respectively) taken from 
Schlegel \etal (1998), using the empirical selective extinction 
function of Cardelli \etal (1989), with $R_V=A_V/E(B-V)=3.1$. 

In Deep3a-F we selected objects to $\Kv<20$, over a total sky area 
of 320 arcmin$^2$. Simulations of point sources show that in all the 
area the completeness is well above 90\% at these K-band levels.
We recall that objects in Daddi-F were selected to completeness limits
of  $\Kv<18.8$ over an 
area of 700 arcmin$^2$ (but we limit our discussion in this paper to 
the 600 arcmin$^2$ covered by the Subaru observations) and to $\Kv<19.2$ 
over a sub-area of 440 arcmin$^2$ (see Daddi \etal 2000 for more 
details).

The total area surveyed, as discussed in this paper, therefore 
ranges from a combined area of 920 arcmin$^2$ (Daddi-F and 
Deep3a-F) at $\Kv<18.8$ to 320 arcmin$^2$ (Deep3a-F) at $\Kv<20$.

Objects in Deep3a-F and Daddi-F were cross-correlated with those
available from the 2MASS survey (Cutri \etal 2003) in the J and K 
bands, resulting in good photometric agreement at better than the 
3\% level.
In order to further verify the photometric zero points we checked 
the colors of stars (Fig.~\ref{fig:ccd}), selected from the $BzK$ 
diagram following Daddi \etal (2004a; see Sect.\ref{sec:S/G}).
From these color-color planes, we find that the colors of stellar 
objects in our data are consistent with those of Pickles (1998), 
with offsets, if any, of  $< 0.1$ mag at most. 
Similar agreement is found with the Lejeune \etal (1997) models.

\begin{figure*}
\centering 
\includegraphics[angle=-90,width=0.9\textwidth]{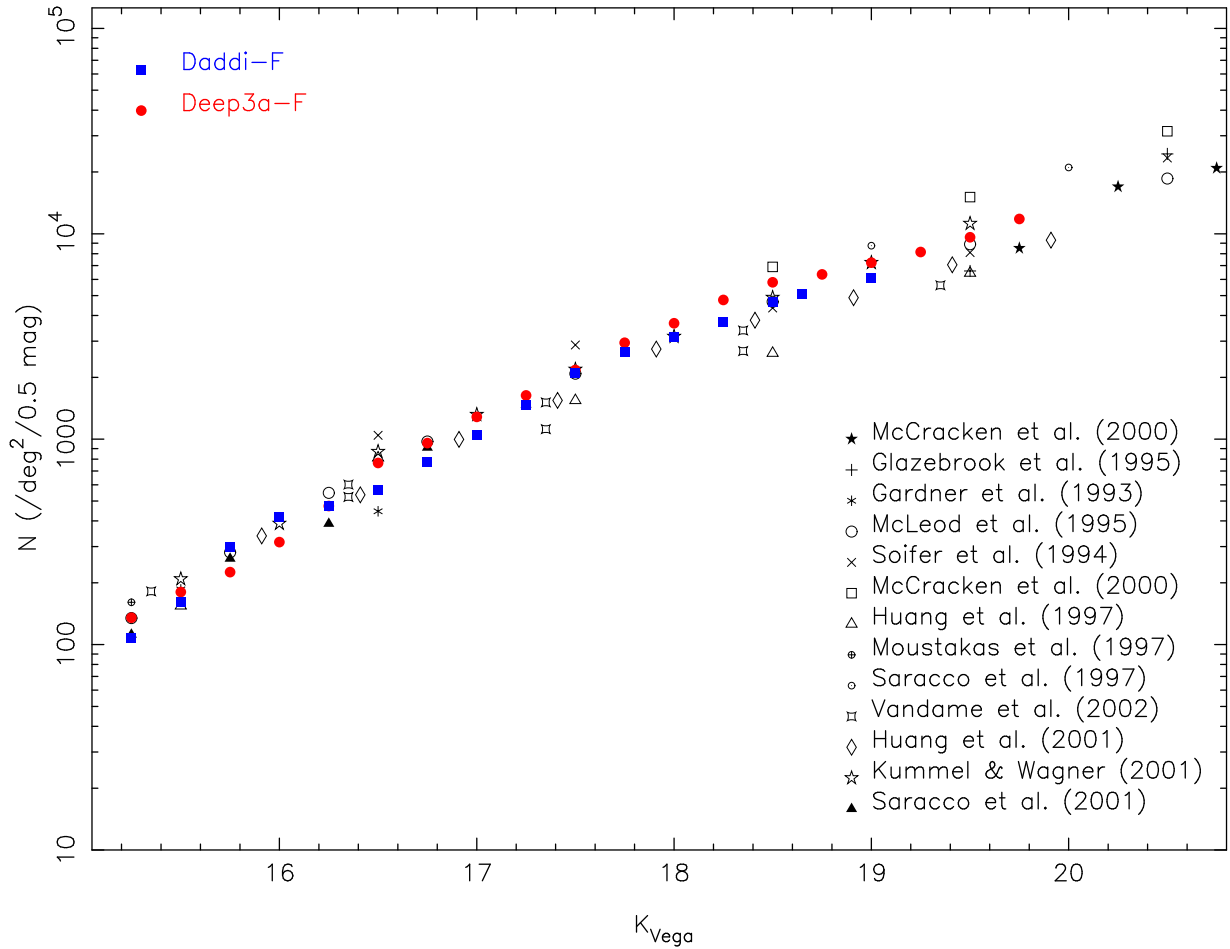}
\caption{
Differential $K$ band galaxy counts from Deep3a-F and Daddi-F, 
compared with a compilation of results taken from various sources.  
[{\it See the electronic edition of the Journal for the color 
version of this figure.}]
}
\label{fig:knumc}
\end{figure*}

Figure~\ref{fig:knumc} shows a comparison of K-band number counts 
in our survey with a compilation of literature counts. No attempt 
was made to correct for different filters ($K_s$ or $K$). 
No corrections for incompleteness were applied to our data, and 
we excluded the stars, using the method in Sect.~\ref{sec:S/G}. 
The filled circles and filled squares correspond, respectively, 
to the counts of Deep3a-F and Daddi-F. As shown in the figure, our 
counts are in good agreement with those of previous surveys.

\subsection{Star-galaxy separation}\label{sec:S/G}

\begin{figure*}
\centering 
\includegraphics[angle=-90,width=0.9\textwidth]{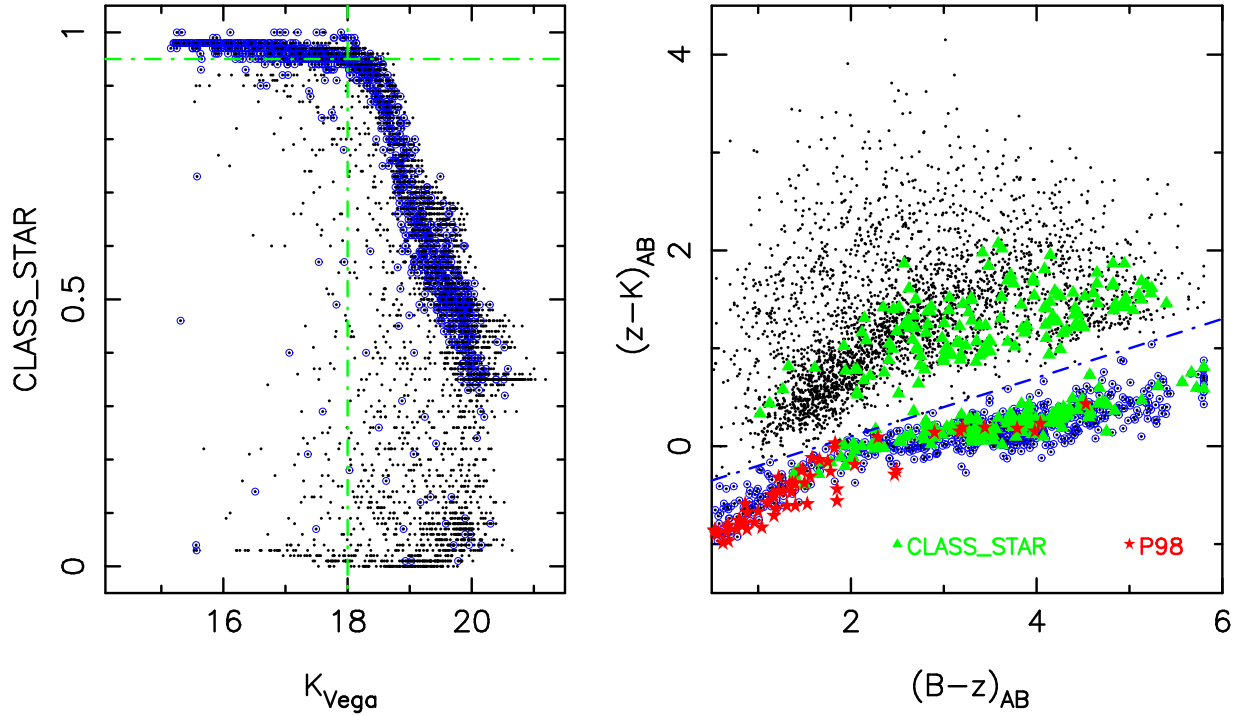}
\caption{
Star/galaxy separation. 
$Left$: K--band magnitude vs. the {\it stellarity index} parameter 
(CLASS\_STAR) from \bsex\ for objects (small dotted points) in 
Deep3a-F.  The dashed lines are ${\rm CLASS\_STAR} = 0.95$ and 
$\Kv =18.0$; objects with $(z-K)_{\rm AB} - 0.3(B-z)_{\rm AB} < 
-0.5$ are plotted as open circles.
$Right$: $B-z$ against $z-K$ for objects in Deep3a-F. objects with 
${\rm CLASS\_STAR} > 0.95$ and $\Kv <18.0$ are plotted as 
triangles, and those with $(z-K)_{\rm AB} - 0.3(B-z)_{\rm AB} < -0.5$ are 
plotted as open circles. Stars correspond to stellar objects 
given in Pickles (1998). 
The dot-dashed line, $(z-K) = 0.3(B-z)-0.5$ , denotes the 
boundary between stars and galaxies adopted in this study; an 
object is regarded as a star, if it is located  below
this line. 
[{\it See the electronic edition of the Journal for the color 
version of this figure.}]
}
\label{fig:SandG}
\end{figure*}

Stellar objects are isolated with the color criterion (Daddi \etal 
2004a) $(z-K)_{\rm AB}<0.3(B-z)_{\rm AB}-0.5$.  In 
Fig.~\ref{fig:SandG} we compare the efficiency of such a color-based 
star-galaxy classification with the one based on the \bsex\ 
parameter CLASS\_STAR, which is based on the shape of the object's 
profile in the imaging data.
It is clear that the color classification is superior, allowing us 
to reliably classify stars up to the faintest limits in the survey. 
However, \bsex\  appear to find resolved in the imaging data a 
small fraction of objects that are color-classified as 
stars. Most likely, these are blue galaxies scattered into the 
stellar color boundaries by photometric uncertainties.

\section{Candidates of 
$\lowercase{\rm s}$B$\lowercase{\rm z}$K$\lowercase{\rm s}$, 
$\lowercase{\rm p}$B$\lowercase{\rm z}$K$\lowercase{\rm s}$ and 
ERO$\lowercase{\rm s}$}\label{sec:cand}

In this section, we select \sbzks, \pegs\ and EROs in the Deep3a-F 
and Daddi-F, using the multicolor catalog based on the NIR K-band 
image (see Sect.~\ref{sec:photo}). 
Forthcoming papers will discuss the selection of DRGs ($J-K>2.3$ 
objects) and LBGs, using the $BRIzJK$ photometry from our database.

\subsection{Selection of \sbzks\ and \pegs}

In order to apply the $BzK$ selection criteria consistently with
Daddi \etal (2004a), we first accounted for the different shapes of
 the filters used, and applied a correction term to the B-band. The
B-band filter used at the Subaru telescope is significantly redder
than that used at the VLT by Daddi \etal (2004a). We then carefully
compared the stellar sequence in our survey to that of Daddi \etal
(2004a), using the Pickles (1998) stellar spectra and the Lejeune 
\etal (1997) corrected models as a guide, and applied small color 
terms to $B-z$ and $z-K$ (smaller than $\sim0.1$ mags in all cases), 
in order to obtain a fully consistent match.
In the following we refer to $BzK$ photometry for the system 
defined in this way, consistent with the original $BzK$ definition by
Daddi \etal (2004a).

\begin{figure*}
\centering 
\includegraphics[angle=-90,width=0.9\textwidth]{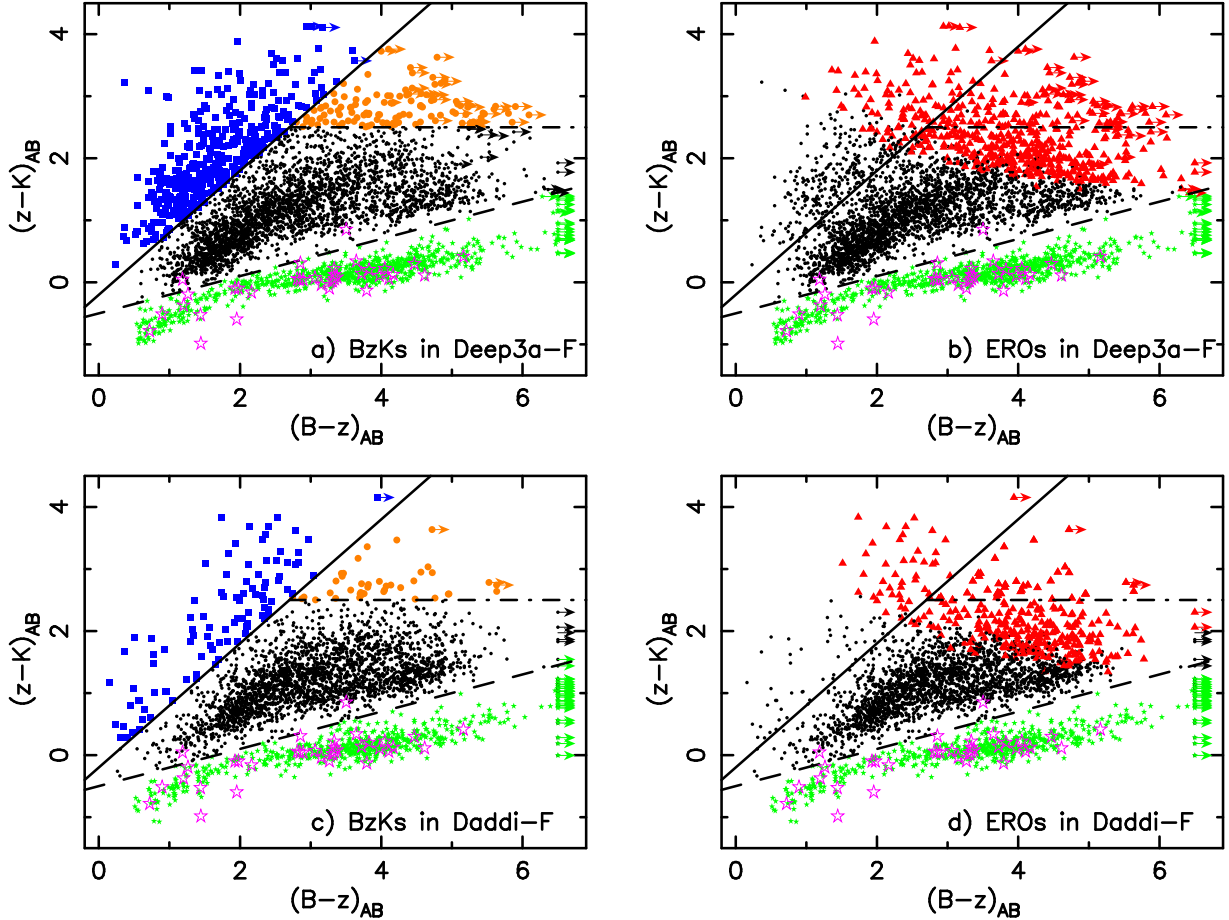}
\caption{
Two-color $(z-K)_{\rm AB}$ vs $(B-z)_{\rm AB}$ diagram for the 
galaxies in the Deep3a-F and Daddi-F fields. Galaxies at 
high redshifts are highlighted. 
The diagonal solid line defines the region 
$BzK\equiv (z-K)_{\rm AB}-(B-z)_{\rm AB}\geq-0.2$ that is efficient 
to isolate $z>1.4$ star forming galaxies (\sbzks). 
The horizontal dot-dashed line further defines the region 
$(z-K)_{\rm AB}>2.5$ that contains old galaxies at $z>1.4$ (\pegs). 
The dashed lines seperate regions occupied by stars and galaxies.
Filled stars show objects classified as stars, having
$(z-K)_{\rm AB} - 0.3(B-z)_{\rm AB} < -0.5$; open stars show
stellar
objects from the K20 survey (Daddi et al. 2005a); 
squares represent \sbzks; circles 
represent \pegs; triangles represent galaxies with 
$(R-K)_{\rm AB}>3.35$ (EROs). Galaxies lying out of the $BzK$ 
regions, thus likely having in general redshifts less than 1.4, are
simply plotted as black points. 
}
\label{fig:bzk}
\end{figure*}

Figure~\ref{fig:bzk} shows  the $BzK$ color diagram of K-selected 
objects in  Deep3a-F and Daddi-F. Using the color criterion from 
Daddi \etal (2004a), $BzK\equiv (z-K)_{\rm AB}-(B-z)_{\rm AB}>-0.2$,
387 galaxies with $K_{\rm Vega} <20$ were 
selected in Deep3a-F as \sbzks, which occupy a narrow
 range on the left of the solid line in Fig.~\ref{fig:bzk}a.  Using 
$BzK< -0.2$ and $(z-K)_{\rm AB}>2.5$, 121 objects 
were selected as candidate \pegs, which lie in the 
top-right part of Fig.~\ref{fig:bzk}a. To $\Kv<20$ a surface density of 
$1.20\pm0.05$ arcmin$^{-2}$\ and of $0.38\pm0.03$ arcmin$^{-2}$ is 
derived separately for \sbzks\ and \pegs, respectively (Poisson's 
errors only). The surface density of \sbzks\ is larger but still
consistent within 2$\sigma$
with the $0.91\pm0.13$~arcmin$^{-2}$ found in the 52 
arcmin$^2$ of the K20 field (Daddi \etal 2004a), and with the 
$1.10\pm0.08$~arcmin$^{-2}$ found in the GOODS North field (Daddi 
\etal 2005b).
Instead, the surface density of \pegs\ recovered here is significantly 
larger than what found in both fields. 
This may well be the result of cosmic variance, given the strong 
clustering of \pegs\ (see Section \ref{sec:clustering}), and their 
lower overall surface density.

Using the same criteria, we select \sbzks\ and \pegs\ in the Daddi-F 
field. In Daddi-F 108 \sbzks\ and 48 \pegs\ are selected, and they 
are plotted in Fig.~\ref{fig:bzk}c. The density of \sbzks\ and \pegs\ 
in Daddi-F is consistent with that in Deep3a, if limited at 
$\Kv\simlt19$.

\subsection{Selection of EROs}
\label{sec:erosel}

EROs were first identified in $K$-band surveys by Elston, Rieke, 
\& Rieke (1988), and are defined here as objects having red 
optical-to-infrared colors such that $(R-K)_{\rm Vega} \ge 5$--6, 
corresponding to $(R-K)_{\rm AB} \ge 3.35$--4.35.
EROs are known to be a mixture of mainly two different populations 
at  $z\gsim0.8$; passively evolving old elliptical galaxies and 
dusty starburst (or edge-on spiral) galaxies whose UV luminosities 
are strongly absorbed by internal dust (Cimatti \etal 2002b; Yan 
\etal 2004, Moustakas \etal 2004). 
In Daddi-F, EROs were selected and studied by Daddi \etal (2000)
using various $R-K$ thresholds.
In order to apply a consistent ERO selection in Deep3a-F, we 
considered the filter shapes and transmission curves. While the same 
K-band filters (and the same telescope and instrument) were used for 
K-band imaging in the two fields, the R-band filters used in the two 
fields 
differ substantially. In the Daddi-F the WHT R-band filter was used,
which is very similar, e.g., to the R-band filter of FORS at 
the VLT used by the K20 survey (Cimatti \etal 2002a). The 
Subaru+Suprime-Cam $R$-band filter is much narrower than the above, 
although it does have a very close effective wavelength. As a result, 
distant $z\sim1$ 
early-type galaxy spectra as well as M-type stars appear to have much 
redder $R-K$ color, by about $\approx$0.3 mag, depending on exact 
redshift and spectral shape.
Therefore, we selected EROs in Deep3a-F with the criterion 
$R_{Subaru}-K>3.7$ (AB magnitudes), corresponding closely to 
$R_{WHT}-K>3.35$ (AB magnitudes) or $R_{WHT}-K>5$ (Vega magnitudes).
In Deep3a-F, 513 EROs were selected to $\Kv<20$, and they are plotted in 
Fig.~\ref{fig:bzk}b with solid red triangles, for a surface density
of 1.6~arcmin$^{-2}$. To the same $\Kv<20$ limit, this agrees well with 
the density found, e.g., in the K20 survey ($\sim1.5$~arcmin$^{-2}$), 
or in the 180 arcmin$^2$ survey by Georgakakis \etal (2005).
In the Daddi-F, 337 EROs were selected with the criterion
$R_{WHT}-K>3.35$, consistent with what done in Deep3a, and are 
plotted in Fig.~\ref{fig:bzk}d as red solid triangles. The surface 
density of EROs in both fields at $\Kv<18.4$ can be compared, with 
overall good consistency, to the one derived from the large 1~deg$^2$ 
survey by Brown \etal (2005).

The peak of the EROs redshift distribution is at $z\sim1$ (e.g., 
Cimatti \etal 2002a).
By looking at the $BzK$ properties of EROs we can estimate how many 
of them lie in the high-$z$ tail $z>1.4$, thus testing the shape of 
their redshift distribution. 
In the Deep3a-F to $\Kv<20$ some 90 of the EROs are also \sbzks, thus 
likely belong to the category of dusty starburst EROs at $z>1.4$, 
while 121 EROs are classified as \pegs. 
Totally, $\sim41$\% of EROs are selected with the $BzK$ criteria, 
thus expected to lie in the high-z tail ($z>1.4$) for $\Kv<20$ 
sample. 
This result is consistent with the value of 35\% found in the 52 
arcmin$^2$ of the K20 field (Daddi \etal 2004a), and with the 
similar estimates of Moustakas \etal (2004) for the full 
GOODS-South area.
In the Daddi-F, to $\Kv<19.2$, 49 of the EROs are also \sbzks; and 
48 of them are also \pegs. 
About 29\% of EROs at $\Kv<19.2$ are in the high-z tail at $z>1.4$.

\subsection{Number counts of EROs, \sbzks, and \pegs}

\begin{table*}
\caption{Differential number counts in 0.5 magnitude bins of 
EROs, \sbzks, and \pegs\ in Deep3a-F and Daddi-F.}\label{tab:numc_highz}
\centering
\scriptsize
%\begin{tabular*}{0.9\textwidth}{@{\extracolsep{\fill}}lccccrrcccc}
\begin{tabular*}{1.0\textwidth}{@{\extracolsep{\fill}}lccccrcccc}
\tableline
\tableline
\multicolumn{5}{c}{Deep3a-F in log (N/deg$^2$/0.5mag)}&
\multicolumn{5}{c}{Daddi-F in log (N/deg$^2$/0.5mag)}\\
\cline{1-5} \cline{6-10} 
K bin center&Galaxies&EROs    & \sbzks & \pegs  &  
K bin center&Galaxies&EROs    & \sbzks & \pegs  \\
\tableline
16.75 &  2.981& 1.353&   ---&   ---& 16.75 &  2.888 &  1.254 &    --- &    ---\\
17.00 &  3.109& 1.654&   ---&   ---& 17.00 &  3.023 &  1.555 &    --- &    ---\\
17.25 &  3.213& 1.830&   ---&   ---& 17.25 &  3.165 &  1.891 &    --- &    ---\\
17.50 &  3.337& 2.198&   ---&   ---& 17.50 &  3.323 &  2.120 &    --- &    ---\\
17.75 &  3.470& 2.483& 1.353&   ---& 17.75 &  3.426 &  2.321 &  1.254 &    ---\\
18.00 &  3.565& 2.675& 1.654& 1.052& 18.00 &  3.498 &  2.590 &  1.622 &  1.078\\
18.25 &  3.678& 2.822& 2.006& 1.830& 18.25 &  3.569 &  2.786 &  1.923 &  1.379\\
18.50 &  3.764& 3.025& 2.228& 2.353& 18.50 &  3.669 &  2.888 &  2.209 &  1.891\\
18.75 &  3.802& 3.138& 2.529& 2.467& 18.65 &  3.708 &  2.990 &  2.342 &  2.175\\
19.00 &  3.859& 3.145& 2.724& 2.596& 19.00 &  3.783 &  3.063 &  2.834 &  2.461\\
19.25 &  3.911& 3.162& 2.971& 2.608&   --- &    --- &    --- &    --- &    ---\\
19.50 &  3.983& 3.201& 3.228& 2.675&   --- &    --- &    --- &    --- &    ---\\
19.75 &  4.072& 3.297& 3.470& 2.759&   --- &    --- &    --- &    --- &    ---\\
\tableline
\end{tabular*}
\end{table*}

\begin{figure*}
\centering 
\includegraphics[angle=-90,width=0.9\textwidth]{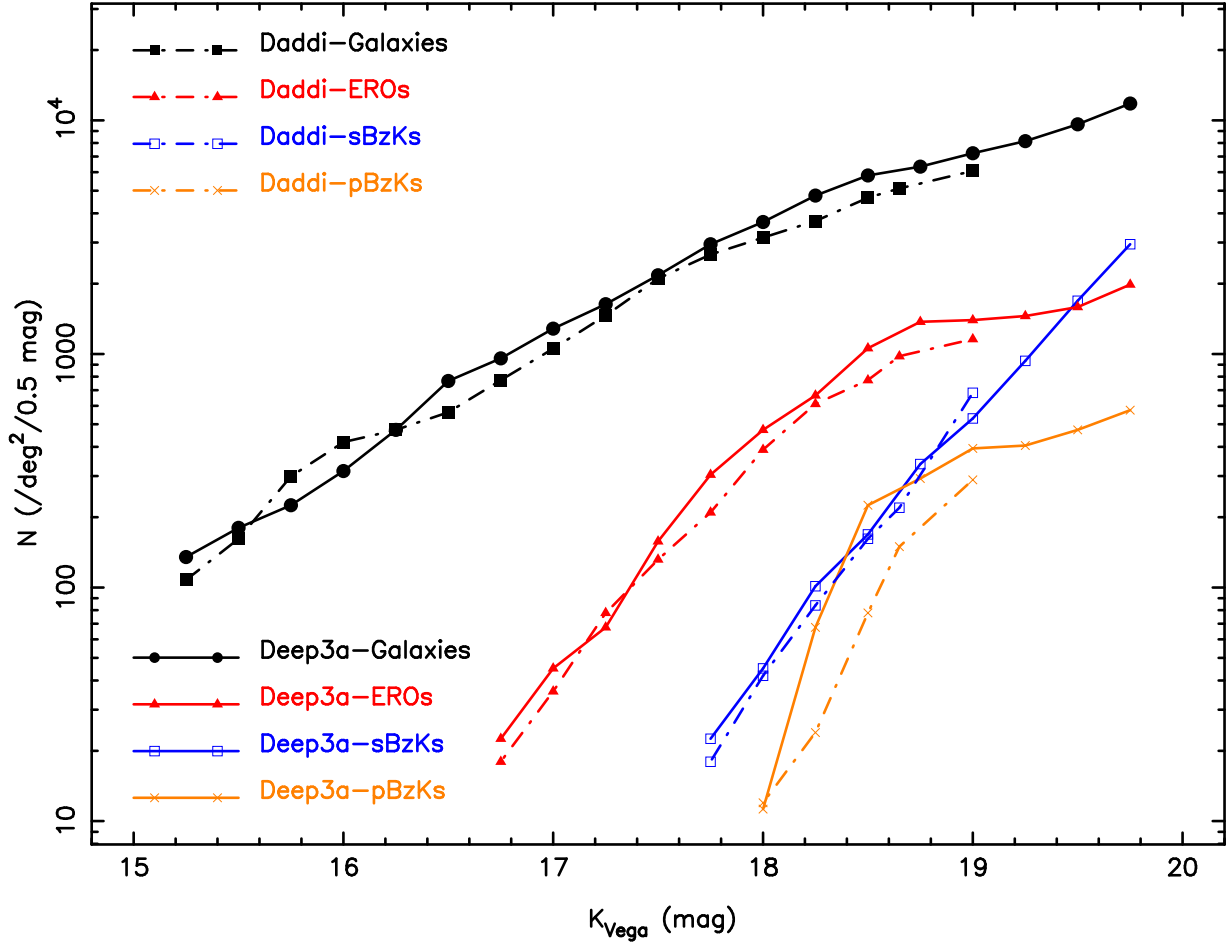}
\caption{K-band differential galaxy number counts for EROs, \sbzks\ 
and \pegs, compared with the $K$-selected field galaxies as in 
Fig.~\ref{fig:knumc}.
The solid curves show the number counts for objects in Deep3a-F, and
the dot-dashed curves show the number counts for objects in Daddi-F.
Triangles, open squares and crosses show the number counts for EROs, 
\sbzks\ and \pegs, respectively. The circles and squares show the 
$K$-selected field galaxies in Deep3a-F and Daddi-F, respectively.
}
\label{fig:numcz}
\end{figure*}

Simple surface densities provide limited insight into the nature of 
different kind of galaxies. However, number-magnitude relations, 
commonly called number counts, provide a statistical probe of both
the space distribution of galaxies and its evolution. For this 
reason, we derived $K$-band differential number counts for EROs,
\sbzks\ and \pegs\ in our fields, and plotted them in 
Figure~\ref{fig:numcz}. 
The differential number counts in 0.5 mag bins are shown in 
 Table~\ref{tab:numc_highz}. Also shown are the number counts for 
all $K$-selected field galaxies in Deep3a-F (circles with solid line) 
and Daddi-F (squares with dot-dashed line), with the same as in 
Fig.~\ref{fig:knumc} for comparison. 
A distinguishable characteristic in Fig.~\ref{fig:numcz} is 
that all of high redshift galaxies (EROs, \sbzks, and \pegs) have
faint NIR apparent magnitudes ($K_{\rm Vega}\gsim$17 mag), and the
slopes of the counts for EROs, \sbzks\ and \pegs\ are steeper than
that of the full $K$-selected sample.

The open squares with a solid line in Fig.~\ref{fig:numcz} shows the 
number counts for \sbzks\ in Deep3a-F. The fraction of \sbzks\ in 
Deep3a-F increases very steeply towards fainter magnitudes. 
The triangles with a solid line and crosses with a solid line show, 
respectively, the number counts of EROs and \pegs\ in Deep3a-F. 
The open squares with a dot-dashed line in Fig.~\ref{fig:numcz} show 
the number counts for \sbzks\ in Daddi-F. The counts of \sbzks\ in 
Daddi-F are almost identical to those in Deep3a-F, to their limit of 
$\Kv\sim19$. The triangles with a dot-dashed line and crosses with 
a dot-dashed line in Fig.~\ref{fig:numcz} show respectively the number 
counts for EROs and \pegs\ in Daddi-F. 

For EROs, the slope of the number counts is variable, being steeper 
at bright magnitudes and flattening out toward faint magnitudes. A 
break in the counts is present at $\Kv\sim 18.0$, very similar to 
the break in the ERO number counts observed by McCarthy \etal (2001) 
and Smith \etal (2002). The \pegs\ number counts have a similar 
shape, but the break in the counts slope is apparently shifted  
$\sim1$--1.5 mag fainter.
There are indications that
EROs and \pegs\ have fairly narrow redshift distributions: 
peaked at $z\sim1$ for EROs (Cimatti \etal 2002b; Yan \etal 2004; 
Doherty \etal 2005) and at $z\approx1.7$ for \pegs\ (Daddi \etal 2004b; 
2005a). The number counts might therefore be direct probes of their 
respective luminosity function, and the shift in the counts is  indeed
consistent with the different typical redshifts of the two population 
of galaxies.

The counts of \sbzks\ have roughly the same slope at all K-band 
magnitudes. This is consistent with the much wider redshift 
distribution of this class of galaxies.
However, we expect that at bright magnitudes AGN contamination might 
be more relevant than at $\Kv\sim20$. Correcting for this, the counts 
of non-AGN \sbzks\ galaxies at bright magnitudes might be 
intrinsically steeper.

\section{Clustering of K-selected galaxy populations} 
\label{sec:clustering}

Measuring the clustering of galaxies provides an additional tool for 
studying the evolution of galaxies and the formation of structures. 
In this section we estimate over the two fields the angular 
correlation of the general galaxy population as well as of the EROs, 
\sbzks\ and \pegs.

In order to measure the angular correlation function of the various 
galaxy samples, we apply the Landy \& Szalay technique (Landy 
\& Szalay 1993; Kerscher \etal 2000), following the approach already 
described in Daddi \etal (2000), to which we refer for formulae
and definitions.

\subsection{Clustering of the $K$-selected field galaxies}

In our analysis a fixed slope $\delta=$\ 0.8 was assumed for the 
two-point correlation function [$w(\theta)=A\times10^{-\delta}$]. 
This is consistent with the typical slopes measured in both faint and 
bright surveys and furthermore makes it possible to directly 
compare our results with the published ones that were obtained 
adopting the same slope (see Daddi \etal 2000 for more details).

\begin{figure*}
\centering 
\includegraphics[angle=0,width=0.75\textwidth]{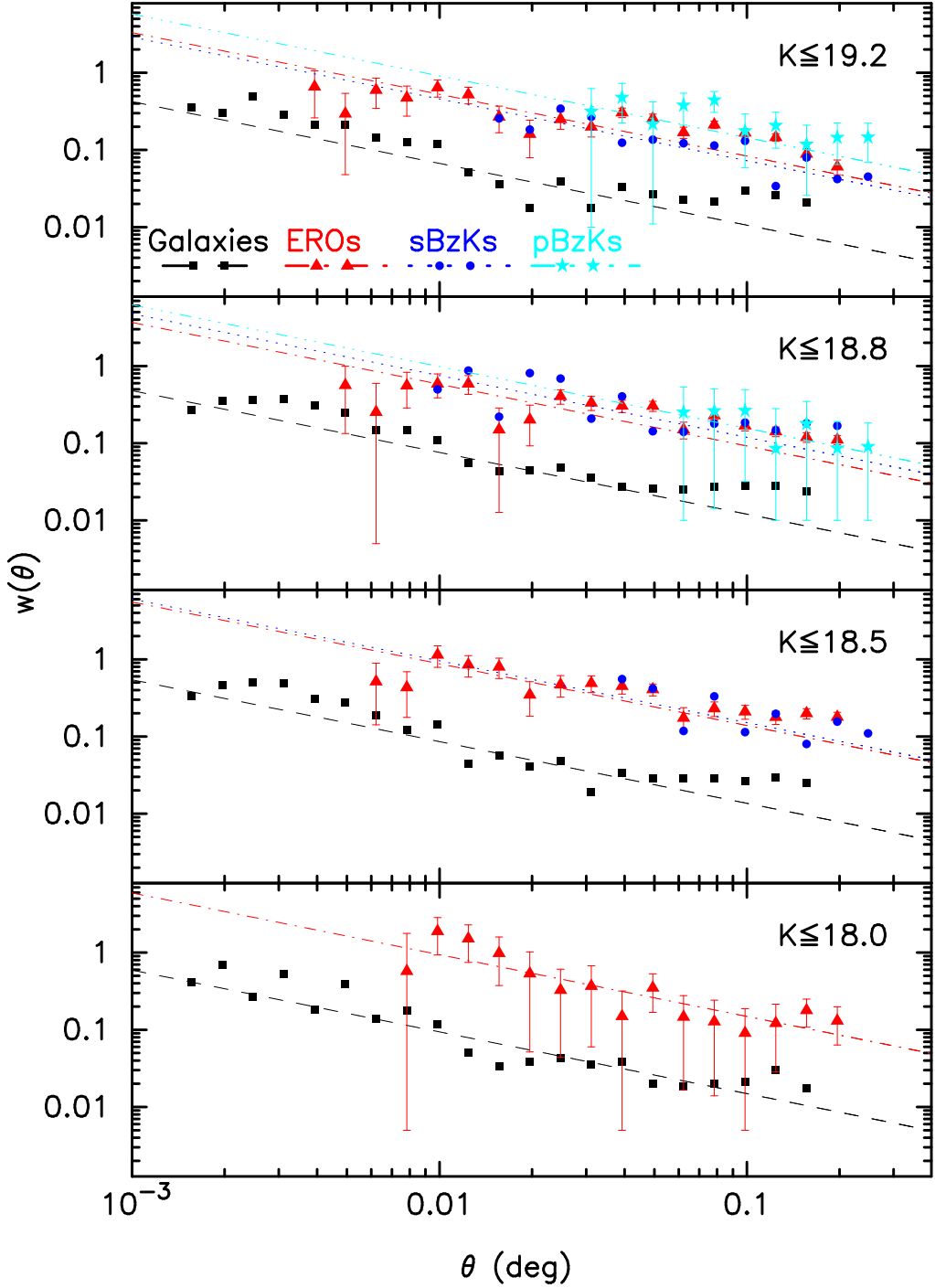}
\caption{
Observed, bias-corrected two-point correlations for the 
Daddi-F sample of field galaxies (squares), EROs (triangles), 
\sbzks\ (filled circles), and \pegs\ (stars).
The error bars on the direct estimator values are $1 \sigma$
errors. 
To make this plot clear, we show the error bars of EROs and 
\pegs\ only, the error bars of field galaxies are smaller, 
and the error bars of \sbzks\ are larger than those of EROs.
The lines (dashed, dot-dashed, dotted, dash-dot-dotted) show 
the power-law fitted to the $w(\theta)$.
Because of the small number of objects included, some bins 
were not populated. 
}
\label{fig:dad_clu}
\end{figure*}

\begin{figure*}
\centering 
\includegraphics[angle=0,width=0.75\textwidth]{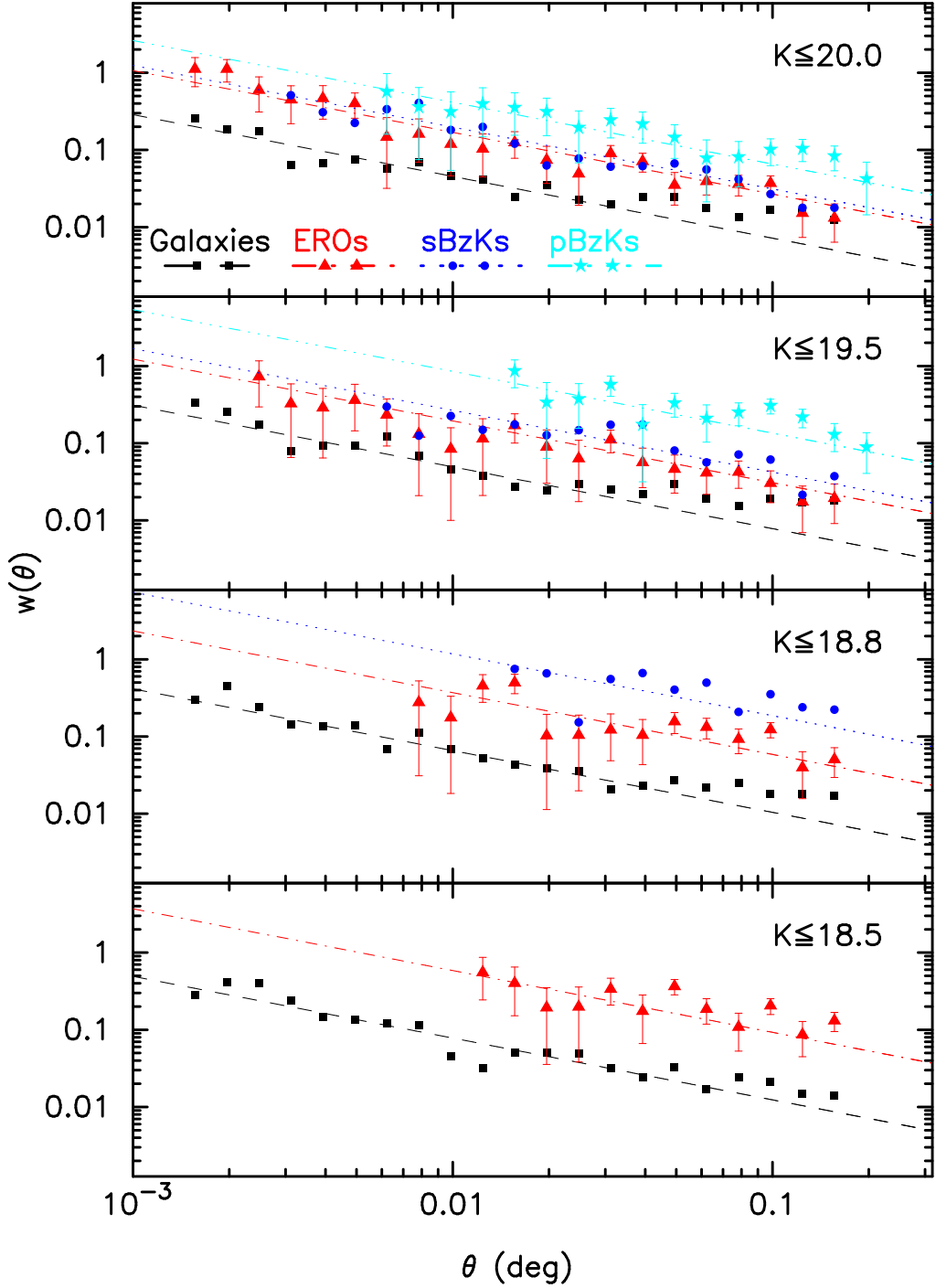}
\caption{Same as Fig.\ref{fig:dad_clu}, but for Deep3a-F. 
}
\label{fig:d3a_clu}
\end{figure*}

In Figures~\ref{fig:dad_clu} and \ref{fig:d3a_clu} the 
bias-corrected two-point correlation functions $w(\theta)$ of Daddi-F 
and Deep3a-F are shown as squares; the bins have a constant 
logarithmic width ($\Delta\log\theta=0.1$), with the bin centers 
ranging from 5.6$\arcsec$ to 11.8$\arcmin$ for both fields. 
These values of $\theta$ 
are large enough to avoid problems of under-counting caused by the 
crowding of galaxy isophotes and yet are much smaller than the extent 
of the individual fields. 
The dashed line shows the power-law correlation function given by a 
least square fit to the measured correlations. We clearly detect a 
positive correlation signal for both fields with an angular 
dependence broadly consistent with the adopted slope $\delta=$\ 0.8.
The derived clustering amplitudes (where $A$ is the amplitude of the 
true angular correlation at 1$^{\circ}$) are presented in the third 
column of Table~\ref{tab:dad_clu} and of Table~\ref{tab:d3a_clu} for 
Daddi-F and Deep3a-F, respectively, and shown in 
Fig.~\ref{fig:angular}.
We compare our result on Daddi-F with those previously 
reported by Daddi \etal (2000), since we use the same $K$-band images 
for this field. The present $K$-selected galaxy sample is however 
different from that of Daddi \etal (2000) in thta: 
(1) the area  (600 $arcmin^2$) is smaller than that (701 $arcmin^2$) 
in Daddi \etal (2000), because we limit ours to the area with all 
the $BRIzK$-band data; and
(2) the star-galaxy separation was done by different methods. In
Daddi \etal (2000) the \bsex\, CLASS\_STAR {\it morphological}
parameter was used, while the {\it photometric} criterion of Daddi 
\etal (2004a) is used here.
Column 10 of Table~\ref{tab:dad_clu} lists the clustering amplitudes 
of all the $K$-selected galaxies in Daddi \etal (2000). 
For $K_{\rm Vega}=18.5$ and fainter bins the $A$-values in the two 
samples are in fair agreement, but for the brightest bin 
($K_{\rm Vega}=18$) the $A$-value in Daddi \etal (2000) is 
$1.3\times10^{-3}$, 45\% smaller than that found here 
($2.36\times10^{-3}$), probably mainly due to the more efficient 
star-galaxy separation employed here (Sect.~\ref{sec:S/G}).
Apart from this small discrepancy, we find good agreement between 
our results and those of Daddi \etal (2000) and Oliver \etal (2004).

The clustering amplitudes in Deep3a-F tend to be slightly but 
systematically smaller than in Daddi-F, which is likely caused 
by the intrinsic variance among different fields, depending on the 
survey geometry, surface density and clustering properties (see 
Sect.~\ref{sec:cos}).

\begin{table*}
\centering
\caption{Clustering amplitudes for the K-selected sample,  
EROs, \sbzks, and \pegs\ in Daddi-F.}\label{tab:dad_clu}
\scriptsize
\begin{tabular*}{1.0\textwidth}{@{\extracolsep{\fill}}lrrrrrrrrrr}
\tableline
\tableline
        & \multicolumn{2}{c}{Galaxies}&\multicolumn{2}{c}{EROs} &
          \multicolumn{2}{c}{\sbzks}  &\multicolumn{2}{c}{\pegs}&
          \multicolumn{2}{c}{Daddi \etal (2000)}\\
  \cline{2-3} \cline{4-5} \cline{6-7}  \cline{8-9} \cline{10-11}
K limit\tablenotemark{a} &    num.  &A[10$^{-3}$]\tablenotemark{b} &   
num.   &A[10$^{-3}$]&     num.  &A[10$^{-3}$]&      num.  &A[10$^{-3}$]  
& Gal.\tablenotemark{c} &EROs\tablenotemark{c}\\	
\tableline
   18.0 &   978& 2.36$\pm$0.94&   51&23.60$\pm$4.18&    7&           ---&    1&           ---&1.3&24\\
   18.5 &  1589& 2.16$\pm$0.40&  132&22.00$\pm$2.82&   21&24.00$\pm$9.80&    5&           ---&1.6&22\\
   18.8 &  2089& 1.91$\pm$0.30&  228&14.60$\pm$1.64&   43&18.90$\pm$7.52&   20&24.70$\pm$9.92&1.5&14\\
   19.2 &  2081& 1.68$\pm$0.28&  264&13.20$\pm$1.26&   92&11.50$\pm$6.61&   40&22.90$\pm$7.63&1.6&13\\
\tableline
\end{tabular*}
\tablenotetext{a}{The area for $K_{Vega}\leq18.8$ mag is about 600 
arcmin$^2$, and for $K_{Vega}=19.2$ it is about 440 arcmin$^2$.}
\tablenotetext{b}{The amplitude of the true angular correlation at 
1$^{\circ}$, the value of the $C$ is 5.46 and 5.74 for the 
whole and the deeper area, respectively.}
\tablenotetext{c}{The last two columns show the clustering 
amplitudes for the K-selected galaxies and the EROs in Daddi 
et al. (2000).}
\end{table*}

\begin{table*}
\centering
\caption{Clustering amplitudes for the K-selected sample,  
EROs, \sbzks, and \pegs\ in Deep3a-F.}\label{tab:d3a_clu}
\scriptsize
\begin{tabular*}{1.0\textwidth}{@{\extracolsep{\fill}}lrrrrrrrrrrr}
\tableline
\tableline
       &\multicolumn{2}{c}{Galaxies}& &\multicolumn{2}{c}{EROs}   & &\multicolumn{2}{c}{\sbzks}&&\multicolumn{2}{c}{\pegs}\\
\cline{2-3} \cline{5-6} \cline{8-9} \cline{11-12} 
K limit &    num.  &A[10$^{-3}$]\tablenotemark{a}& &   num.   &A[10$^{-3}$]& &    num.  &A[10$^{-3}$]& &    num.  &A[10$^{-3}$]\\	
\tableline
   18.5 &   997& 1.96$\pm$1.04&&   95&14.70$\pm$2.20&&   13&           ---&&    8&           ---\\
   18.8 &  1332& 1.65$\pm$0.76&&  166& 9.29$\pm$1.60&&   27&29.50$\pm$5.80&&   26&40.90$\pm$7.55\\
   19.5 &  2284& 1.24$\pm$0.41&&  340& 4.89$\pm$0.78&&  129& 6.70$\pm$3.14&&   71&21.40$\pm$3.00\\
   20.0 &  3333& 1.14$\pm$0.28&&  513& 4.25$\pm$0.52&&  387& 4.95$\pm$1.69&&  121&10.40$\pm$2.83\\
\tableline
\end{tabular*}
\tablenotetext{a}{The amplitude of the true angular correlation at 
1$^{\circ}$, the value of the $C$ is 6.63.}
\end{table*}

We find a smooth decline in amplitude with K-band magnitude that is 
consistent with the results from Roche \etal (2003), Fang \etal 
(2004), and Oliver \etal (2004). The decline is not as steep as in 
the range $15<\Kv<18$ (see e.g. Roche \etal 1999). The flattening, 
which extends up to $K\sim22$--24 (Carlberg \etal 1997; Daddi \etal 
2003) has been interpreted as due to the existence of strongly 
clustered K-selected galaxy populations extending to redshifts 
$z\sim1$--3 (Daddi \etal 2003).
Beside the well known 
$z\approx1$ EROs strongly clustered galaxy populations, other 
populations with high angular clustering indeed exist with redshift 
extending to $z\approx2.5$ at least, as discussed in the next sections.

\subsection{Clustering of the EROs}

We estimate the clustering properties of the EROs, using the large 
sample of EROs derived from our two fields.
Figures~\ref{fig:distrib_obj}a) and \ref{fig:distrib_obj}b)  clearly 
show, from both fields, that the sky distribution of EROs is very 
inhomogeneous, as first noted by Daddi \etal (2000).
The two-point correlation functions are shown in 
Fig.~\ref{fig:dad_clu} and \ref{fig:d3a_clu} for Daddi-F and 
Deep3a-F, and the dot-dashed lines show the power-law correlation 
function given by a least squares fit to the measured values. The 
correlations are well fitted by a $\delta=0.8$ power law. 

\begin{figure*}
\centering 
\includegraphics[angle=-90,width=0.9\textwidth]{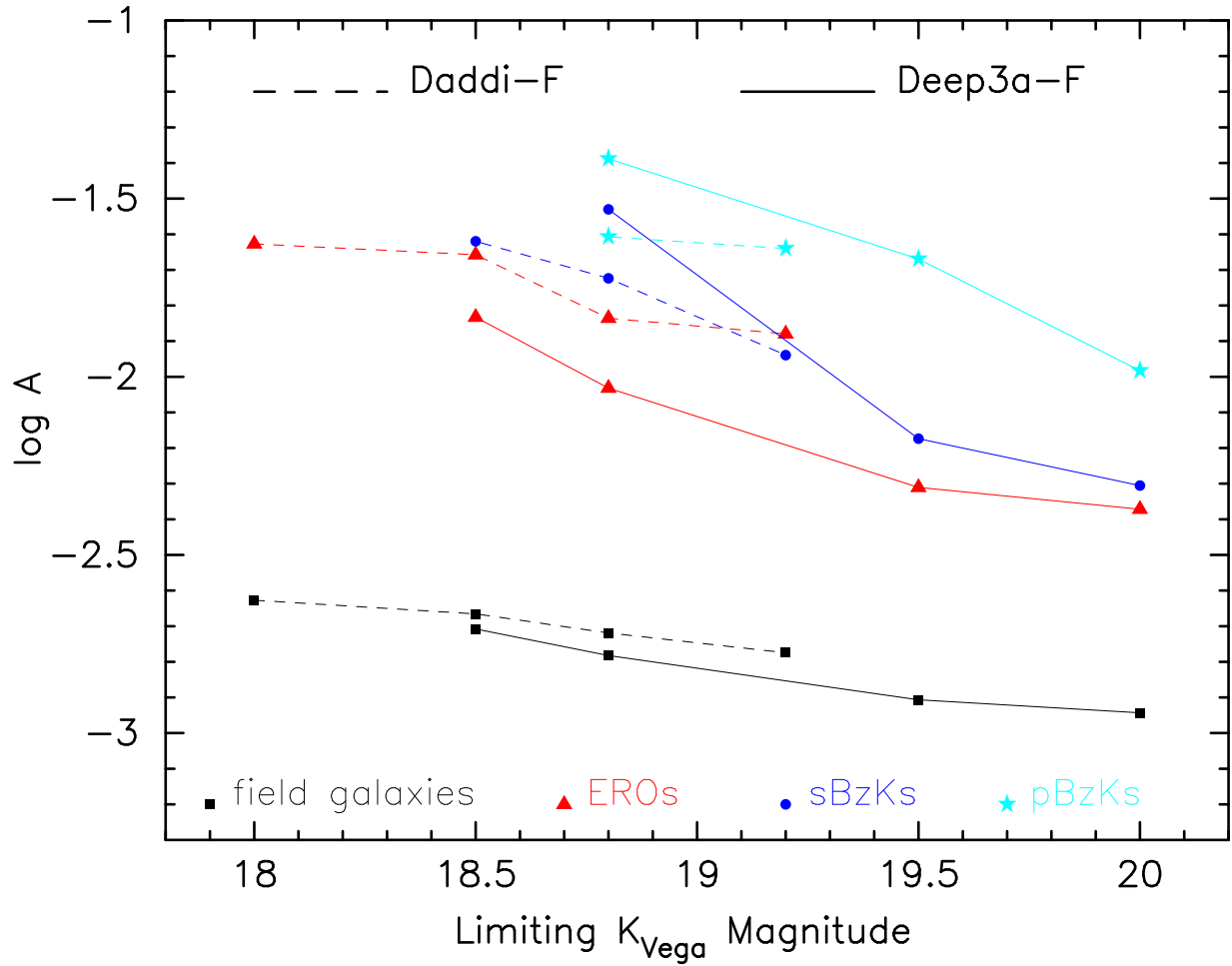}
\caption{Angular clustering amplitudes of field galaxies, EROs,
\sbzks, and \pegs\ shown as a function of the $K$-band limiting 
magnitudes of the sample analyzed. Tables~\ref{tab:dad_clu} and 
\ref{tab:d3a_clu} summarize the measurements together with their 
(Poisson) errors.
[{\it See the electronic edition of the Journal for the color version 
of this figure.}]
}
\label{fig:angular}
\end{figure*}

\begin{figure*} 
\centering 
\includegraphics[width=0.9\textwidth]{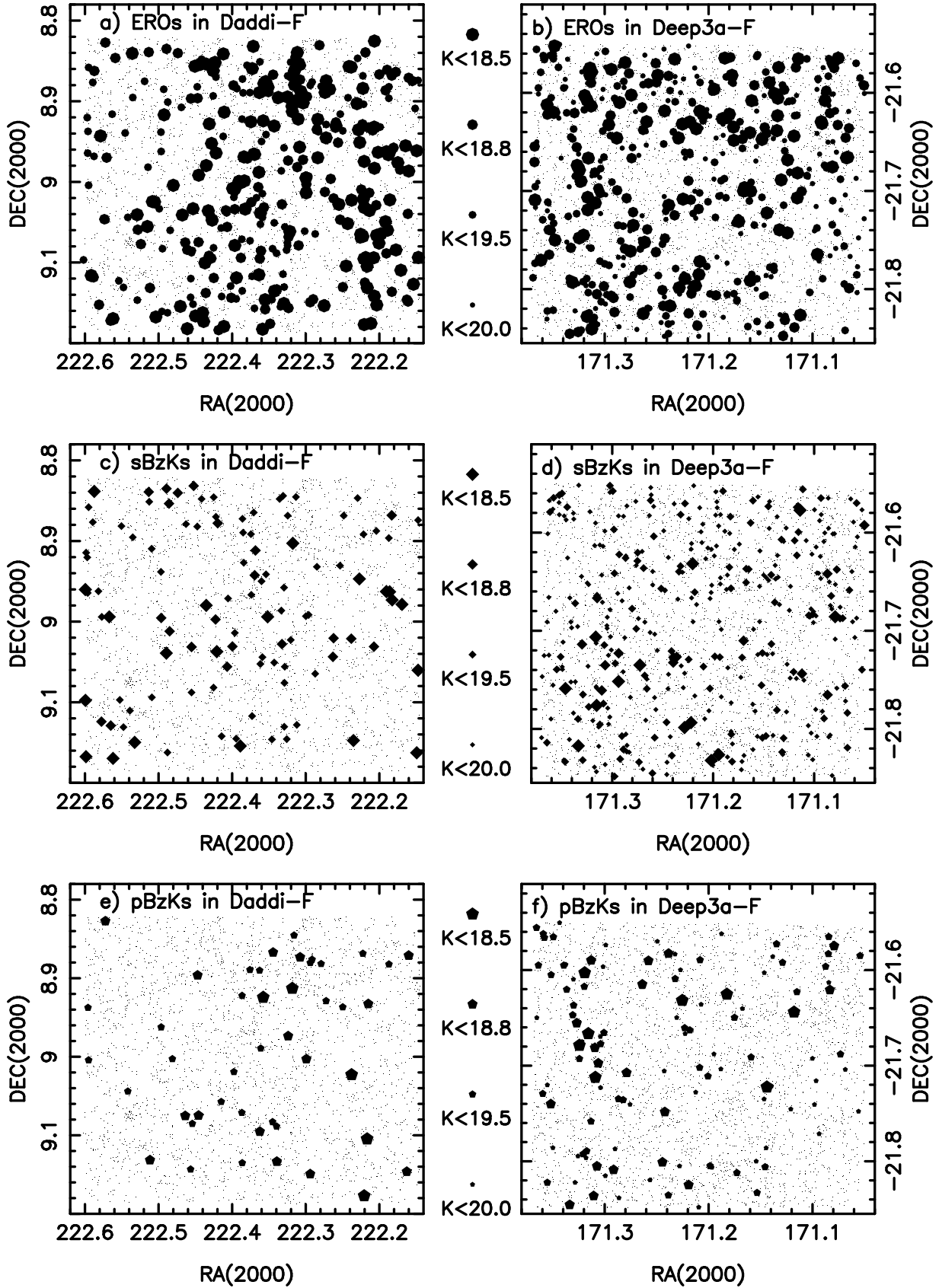}
\caption{
Sky positions of the EROs, \sbzks, and \pegs\ in the Daddi-F 
and Deep3a-F fields. 
(a) EROs in Daddi-F; (b) EROs in Deep3a-F;
(c) \sbzks\ in Daddi-F; (d) \sbzks\ in Deep3a-F;
(e) \pegs\ in Daddi-F; (f) \pegs\ in Deep3a-F.}
\label{fig:distrib_obj}
\end{figure*}

A strong clustering of the EROs is indeed present at all scales that 
could be studied, and its amplitude is about one order of magnitude 
higher than that of the field population at the same $K_{\rm Vega}$ 
limits, in agreement with previous findings (Daddi \etal 2000; Firth 
\etal 2002; Brown \etal 2005; Georgakakis \etal 2005). 
The derived clustering amplitudes are reported in column (5) of 
Table~\ref{tab:dad_clu} and Table~\ref{tab:d3a_clu} for Daddi-F 
and Deep3a-F, respectively. 
The amplitudes shown in Fig.~\ref{fig:angular} suggest a 
trend of decreasing strength of the clustering for fainter EROs in 
both fields.

For the $\Kv < 18.5$ and $18.8$ mag subsamples of Daddi-F, the 
correlation function signal is significant at the 7 $\sigma$ level 
with clustering amplitudes $A = 22.0\times10^{-3}$ and 
$14.6\times10^{-3}$, respectively. For the subsample with 
$\Kv < 19.2$ mag, the detected signal is significant at the 10 
$\sigma$ confidence level ($A = 13.2\times10^{-3}$). 
In column 11 of Table~\ref{tab:dad_clu} we also list the 
clustering amplitudes of EROs in Daddi \etal (2000), which are in 
good agreement with the present findings.
Using a $\sim180$ arcmin$^2$ Ks-band survey of a region within the 
Phoenix Deep Survey, Georgakakis \etal (2005) have analyzed a 
sample of 100 EROs brighter than $\Kv = 19$ mag, and estimated an 
amplitude $A = 11.7\times10^{-3}$, consistent with our results. 

As for the $K$-selected galaxies, the clustering amplitude of EROs 
in Deep3a-F is slightly smaller than that in Daddi-F.  
The clustering amplitudes are $A = 9.29\times10^{-3}$ for $\Kv<18.8$ 
EROs, and $A = 4.25\times10^{-3}$ for $\Kv<20.$ EROs, which is 
weaker than the $A = 8.7\times10^{-3}$ for $\Kv<20$ EROs in 
Georgakakis \etal (2005). 
Field-to-field variation is one of possible reasons for this 
discrepancy, which is discussed in Sect.~\ref{sec:cos}.

We notice that our results are solid against possible contamination
by stars.
Contamination by unclustered populations (i.e. stars) would reduce 
the amplitude of the angular correlation function by $(1-f)^2$, 
where $f$ is the fractional contamination of the sample. The prime 
candidates for contamination among EROs are red foreground Galactic 
stars (note that stars are not contaminating BzK samples), which 
can have red $R-K$ colors. However, we have rejected stars among 
EROs using the photometric criterion for star-galaxy separations 
(see Sect.~\ref{sec:S/G}). Therefore, our clustering measurements 
refer to extragalactic EROs only.

\subsection{Clustering of the star-forming BzKs}

Figures~\ref{fig:distrib_obj}c) and \ref{fig:distrib_obj}d) display 
the sky distribution of the \sbzks\ in Daddi-F and Deep3a-F, and 
shows that also these galaxies have a quite inhomogeneous 
distribution.  
This is not an artifact of variations of the detection limits over 
the fields, because the Monte Carlo simulations show that 
differences in detection completeness within small 
($4\arcmin.0 \times 4\arcmin.0$) areas in the image are very small, 
and that the detection completeness does not correlate with the 
distribution of the \sbzks.

The resulting angular correlation functions for the \sbzks\ are 
shown in Fig.~\ref{fig:dad_clu} and Fig.~\ref{fig:d3a_clu} for the 
two fields. 
Again, a slope $\delta = 0.8$ provides a good fit to the data. The 
best fit values of $A$ are reported in column 7 of 
Table~\ref{tab:dad_clu} and Table~\ref{tab:d3a_clu}.

The $w(\theta=1^{\circ})$ amplitudes of \sbzks\ in Daddi-F are 
$24.0\times10^{-3}$, $18.9\times10^{-3}$ and $11.5\times10^{-3}$
at $\Kv=18.5$, 18.8 and 19.2, respectively. 
For \sbzks\ in Deep3a-F, the $w(\theta)$ amplitudes become 
$29.5\times10^{-3}$, $6.70\times10^{-3}$ and $4.95\times10^{-3}$ 
at $\Kv=18.8$, 19.5 and 20.0, respectively.  
The \sbzks\ appear to be strongly clustered in both fields, and the 
clustering strength increases with the $K$-band flux. 
Actually, they are as strongly clustered as the EROs.
Strong clustering of the \sbzks\ was also inferred by Daddi \etal 
(2004b), by detecting significant redshift spikes in a sample of 
just nine $\Kv<20$ \sbzks\ with spectroscopic redshifts in the range 
$1.7<z<2.3$.
Albeit without spectroscopic redshifts, the present result is 
instead based on a sample of 500 \sbzks. 
Adelberger \etal (2005) also found that UV-selected, star forming 
galaxies with $\Kv<20.5$ in the redshift range $1.8\le z \le 2.6$ 
are strongly clustered. Our results are in good agreement with 
these previous findings.

\subsection{Clustering of the passive BzKs}

Figures~\ref{fig:distrib_obj}e) and \ref{fig:distrib_obj}f) display 
the sky distribution of the \pegs\ in Daddi-F and Deep3a-F, and 
show that also these galaxies have a very inhomogeneous 
distribution. 
We then derive the angular two-point correlation function of \pegs, 
 using the same method as in the previous subsections.
 
The resulting angular correlation functions for the \pegs\ are 
shown in Fig.~\ref{fig:dad_clu} and Fig.~\ref{fig:d3a_clu} for the 
two fields. Again, a slope $\delta = 0.8$ provides a good fit to 
the data. The best fit values of $A$ are reported in column (9) 
of Table~\ref{tab:dad_clu} and Table~\ref{tab:d3a_clu}.

The \pegs\ appear to be the most strongly clustered galaxy 
population in both fields (with $A = 22.9\times10^{-3}$ for 
$\Kv<19.2$ \pegs\ in Daddi-F and $A = 10.4\times10^{-3}$ for 
$\Kv<20.0$ \pegs\ in Deep3a-F), and the clustering strength 
increases with increasing $K$-band flux. 

\subsection{$K$-band dependence and field-to-field variations 
of clustering measurements}\label{sec:cos}

Fig.~\ref{fig:angular} summarizes the clustering measurements for the 
populations examined (field galaxies, EROs, and BzKs), as a function 
of the $K$-band limiting magnitudes of the samples. Clear trends with 
$K$ are present for all samples, showing that fainter galaxies have 
likely lower intrinsic (real space) clustering, consistent with the 
fact that objects with fainter $K$ are less massive, or have wider 
redshift distributions, or both. For the EROs (see Sect.~\ref{sec:erosel})
we have found evidence that faint $\Kv<20$ samples have indeed a higher
proportion of galaxies in the $z>1.4$ tail, with respect to brighter 
$\Kv<19$ objects, and thus a wider redshift distribution.
As already noted, all color-selected high-redshift populations are 
substantially more clustered than field galaxies, at all the 
magnitudes probed here. 
The reason for the stronger angular clustering of \pegs, compared, 
e.g, to \sbzks\ or EROs, is likely (at least in part) their narrower redshift 
distributions $1.4<z<2$ (Daddi \etal 2004a; 2005a). In future papers, 
we will use the Limber equation, with knowledge of the redshift 
distributions, to compare the real space correlation length of the 
different populations.
 
For strongly clustered populations, with angular clustering 
amplitudes $A\approx10^{-2}$, a large cosmic variance of clustering 
is expected, which is relevant also on fields of the size of the ones studied 
here (see Daddi \etal 2001; 2003).
There are in fact variations in our clustering measurements for 
high-redshift objects between the two fields, sometimes larger than 
expected on the basis of the errors on each clustering measurements.
We remind the reader that realistic {\it external} errors on the 
angular clustering of EROs, as well as \sbzks\ and \pegs\ are likely 
larger than the Poissonian ones that we quote. Following the recipes
by Daddi \etal (2001) we estimate that the typical total accuracy for our 
measurements of $A\approx0.02$, when including {\it external} 
variance, is on the order of 30\%.

\subsection{Cosmic variance in the number counts}

The presence of strong clustering will also produce substantial field
to field variations in the number counts. Given the available 
measurements of angular clustering, presented in the previous 
sections, we are able to quantify the expected variance in the 
galaxies counts, following e.g. eq.~(8) of Daddi \etal (2000). We 
estimate that, for the Deep3a-F limit of $\Kv<20$, the integrated 
numbers of objects are measured to $\sim20$\% precision for EROs
and \sbzks, and to 30\% precision for \pegs.

\section{Properties of \sbzks}

The accurate analysis of the physical properties of high redshift
galaxies (such as SFR, stellar mass, etc.) requires the knowledge 
of their spectroscopic redshift. 
VLT VIMOS spectra for objects culled by the 
present sample of \sbzks\ and EROs have been recently secured and 
are now being analyzed, and will be used in future publications. 
In the meantime, estimates of these quantities can be 
derived on the basis of the present photometric data, following the
recipes calibrated in Daddi \etal (2004a), to which we refer  for 
definitions and a more detailed discussion of the recipes.

While errors by a factor of 2 or more may affect individual 
estimates done in this way, when applied to whole population of 
$BzK$-selected galaxies these estimates should be relatively robust, 
on average, because the Daddi et al.'s (2004a) relations were derived 
from a sample of galaxies with spectroscopic redshifts.
The estimates presented here thus represent a significant improvement
 on the similar ones provided by Daddi \etal (2004a) because of the 
6 to 20 times larger area probed (depending on magnitude)
 with respect to the K20 survey, 
which should help to significantly reduce the impact of cosmic 
variance.

\subsection{Reddening and star formation rates}

Following Daddi \etal (2004a), estimates of the reddening $E(B-V)$ 
and SFR for \sbzks\ can be obtained from the $BzK$ colors and fluxes 
alone (see Daddi \etal 2004a for more details). 
The reddening can be estimated by the $B-z$ color, providing a 
measure of the UV slope. The Daddi \etal's recipe is consistent with 
the recipes by Meurer \etal (1999) and Kong \etal (2004) for a 
Calzetti \etal (2000) extinction law, based on the UV continuum slope.
Daddi et al. (2004a) showed that for $BzK$ galaxies this method 
provides $E(B-V)$ with an rms dispersion of the residuals of about 
0.06, if compared to values derived with knowledge of the redshifts, 
and with the use of the full multicolor SED. 
With knowledge of reddening, the reddening corrected B-band flux is 
used to estimate the 1500\AA\ rest-frame luminosity, assuming an 
average redshift of 1.9, which can be translated into SFR on the 
basis, e.g., of the Bruzual \& Charlot (2003) models. 
Daddi et al. (2005b) showed that SFRs derived in this way are 
consistent with radio and far-IR based estimates, for the average 
\sbzks.

The $E(B-V)$ and SFR histograms of the \sbzks\ in Daddi-F and 
Deep3a-F are shown in Fig.~\ref{fig:bzk_nat}. About 95\% of the 
\sbzks\ in Daddi-F ($\Kv < 19.2$) have SFR$>70\,M_\odot$yr$^{-1}$, 
and the median SFR is about $370~M_\odot$yr$^{-1}$.  About 90\% of 
the \sbzks\ in Deep3a-F ($\Kv < 20.0$) have  
SFR$>$70$\,M_\odot$yr$^{-1}$, and the median SFR is 
$\sim190$ $M_\odot$yr$^{-1}$. 

The median reddening for the $\Kv<20$ \sbzks\ is estimated to be 
$E(B-V)=0.44$, consistently with Daddi \etal (2004a; 2005b). Of
\sbzks\ 55\%  have $E(B-V)>0.4$, the limit at which we estimate the 
UV-based criteria of Steidel \etal (2004) would fail at selecting 
$z\sim2$ galaxies. Therefore, we estimate that $\simgt55$\% of the 
$z\sim2$ galaxies would be missed by the UV criteria.
This is similar to, but higher, than the 40\% estimated by Reddy \etal 
(2005).
The probable reason for the small discrepancy is that Reddy \etal 
(2005) excluded from their sample the reddest \sbzks\ for which the 
optical magnitudes could not be accurately measured in their data.

\subsection{Stellar masses of \sbzks\ and \pegs}

Using BC03 models, spectroscopic redshifts of individual K20 
galaxies, and their whole $UBV$$RIz$$JHK$ SED, Fontana \etal 
(2004) have estimated the 
stellar-mass content for all the K20 galaxies (using a Salpeter IMF
from 0.1 to 100 M$_\odot$, as adopted in this paper). The individual 
mass estimates for 31 non-AGN \sbzks\ and \pegs\ objects with 
$z>1.4$ have been used by Daddi \etal (2004a)  to calibrate an 
empirical relation 
giving the stellar mass for both \sbzks\ and \pegs\ as a function 
of their observed $K$-band total magnitude and $z-K$ color. 
This relation allows one to estimate the stellar mass with 
uncertainties on single objects of about 60\% compared to the 
estimates based on knowledge of redshifts and using the full 
multicolor SEDs.
The relatively small variance is introduced by intrinsic 
differences in the luminosity distance, in the $M/L$ ratio for 
given magnitudes and/or colors. 

\begin{figure*}
\centering 
\includegraphics[angle=-90,width=0.9\textwidth]{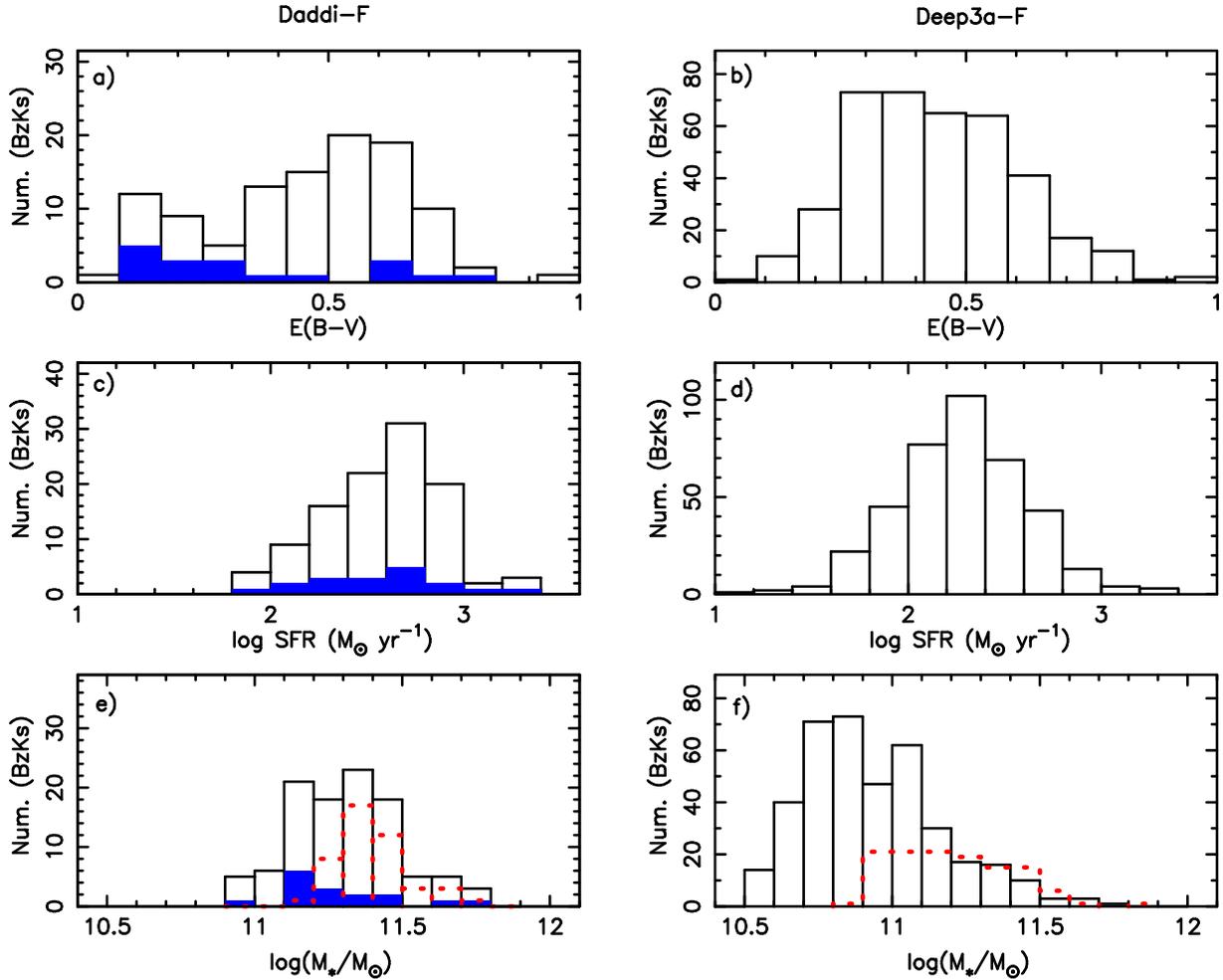}
\caption{
Reddening, star formation rate, and stellar mass histogram 
of \sbzks\ in Daddi-F and Deep3a-F. 
(a), (c), (e) (left panels): Plots for Daddi-F; (b), (d), and (f) 
(right): Plots for Deep3a-F.
The filled area is the histogram for \sbzks\ in Daddi-F, which 
are associated with X-ray sources (about 25\%).
The dashed lines in (e) and (f) are the stellar mass 
histograms of \pegs.
[{\it See the electronic edition of the Journal for the color version 
of this figure.}]
}
\label{fig:bzk_nat}
\end{figure*}

The histograms for the stellar mass of the \sbzks\ derived in 
this way are shown in Fig.~\ref{fig:bzk_nat}e and 
Fig.~\ref{fig:bzk_nat}f.
About 95\% of the \sbzks\ in Daddi-F have $M_*>10^{11} M_\odot$, 
and the median stellar mass is $2.0\times10^{11}M_\odot$; 
in Deep3a-F $\sim 40\%$ of the \sbzks\ have 
$M_*>10^{11} M_\odot$, the median stellar mass 
is $\sim 8.7\times10^{10}M_\odot$, a difference due to the different
limiting $K$ magnitude in the two fields.

Using the same method, we also estimate the stellar mass of the 
\pegs\ in both fields, and plot them in Fig.~\ref{fig:bzk_nat}e 
and Fig.~\ref{fig:bzk_nat}f as the dotted line. The median stellar 
mass of \pegs\ in Daddi-F is $\sim 2.5\times10^{11}M_\odot$, 
and in Deep3a-F is $\sim 1.6\times10^{11}M_\odot$, respectively.
Again, the higher masses for \sbzks\ in Daddi-F compared to 
Deep3a-F result from the shallower $K$-band limit. 

It is worth noting that in the $K_{\rm Vega}<20$ sample there are
barely any \pegs\ less massive than $7\times 10^{10}M_\odot$ (see
Fig.~\ref{fig:bzk_nat}f), while over 50\% of \sbzks\ are less 
massive than this limit. This is primarily a result of \pegs\ being 
by construction redder than $(z-K)_{\rm AB}=2.5$, and hence Eq.~(6) and
Eq.~(7) in  Daddi \etal (2004a)
with $\Kv<20$ implies $M_*\gsim 7\times 10^{10}M_\odot$. Note that 
above $10^{11}M_\odot$ (above which our sample should be reasonably
complete) the numbers of \sbzks\ and \pegs\ are similar. We  
return to this point in the last section.

\subsection{Correlation between physical quantities}

\begin{figure*}
\centering 
\includegraphics[angle=-90,width=0.9\textwidth]{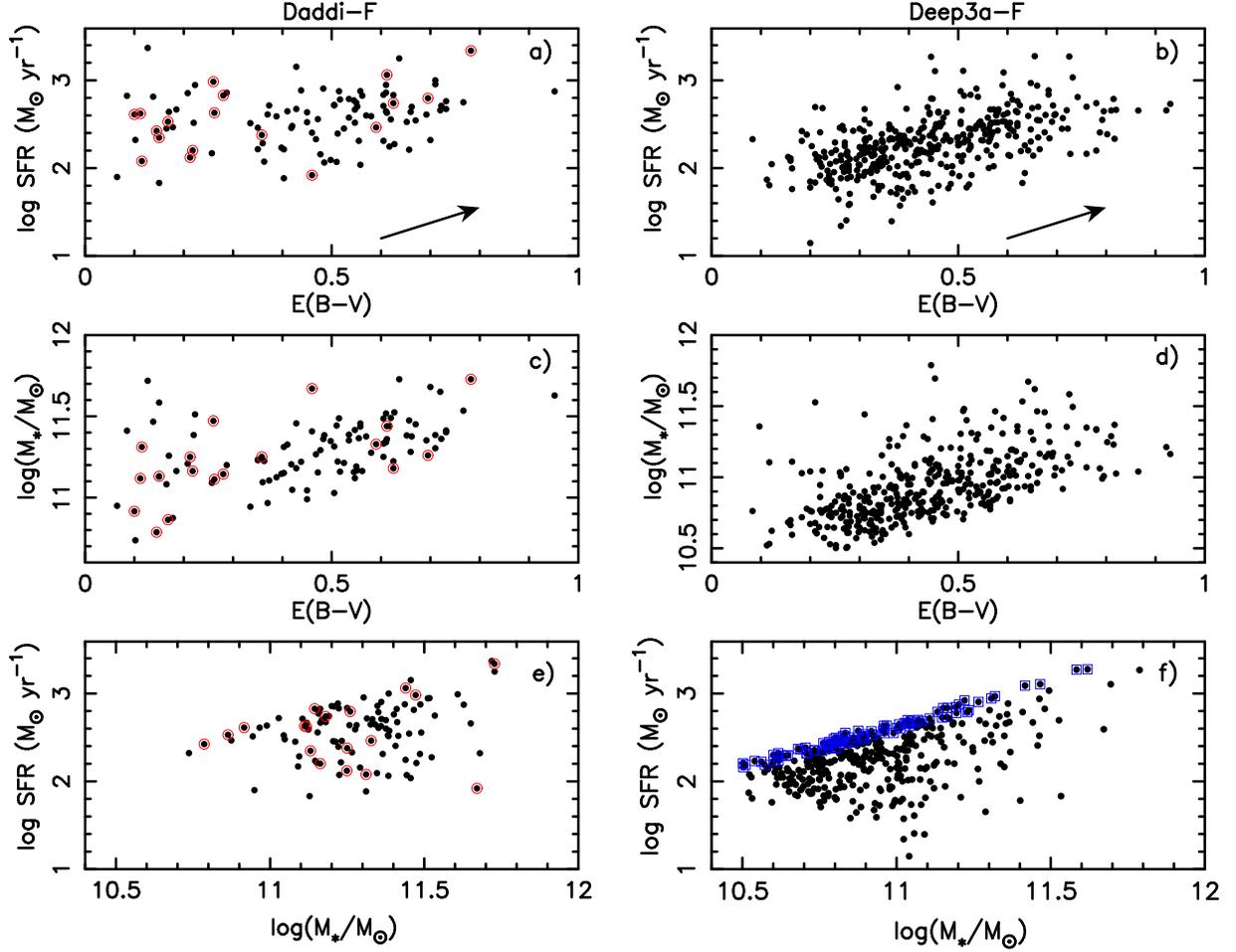}
\caption{Cross correlation plots of the physical quantities estimated 
for \sbzks\ in our fields. Circles are the X-ray detected \sbzks\ in 
Daddi-F.
The arrows in the top panels show the slope of the correlation 
induced by the propagation of the reddening errors into the SFRs.
In the bottom-right panel, squares show objects having $-0.2<BzK<0.0$.
[{\it See the electronic edition of the Journal for the color version 
of this figure.}]}
\label{fig:bzk_cor}
\end{figure*}

Figures~\ref{fig:bzk_cor}a) and \ref{fig:bzk_cor}b) show the 
correlation between color excess $E(B-V)$ and SFR for the \sbzks\ 
in Daddi-F and Deep3a-F, respectively.
The Spearman rank correlation coefficients are $r_s=0.40$ for 
Daddi-F and $r_s=0.57$ for Deep3a-F. This implies that the SFR is 
significantly correlated to $E(B-V)$, at a $>5\,\sigma$ level, and 
the reddest galaxies have the highest SFR.
Part of this correlation can arise from simple error propagation, 
as an overestimate (underestimate) of the reddening automatically 
leads to an overestimate (underestimate) of the SFR of a galaxy. 
In Fig.~\ref{fig:bzk_cor}a) and Fig.~\ref{fig:bzk_cor}b) the arrow 
shows the resulting slope of the correlated reddening errors, 
which indeed is parallel to the apparent correlation. However, 
the scatter in the original SFR-$E(B-V)$ correlation 
[$\delta E(B-V)\sim 0.06$; Daddi \etal\ 2004a] is much smaller 
than what needed to produce the full correlation seen in 
Fig.~\ref{fig:bzk_cor}a) and \ref{fig:bzk_cor}b).
We conclude that there is evidence for an {\it intrinsic} 
correlation between SFR and reddening for $z\sim 2$ star-forming 
galaxies, with galaxies with higher star formation rates having
more dust obscuration.
A positive correlation between SFR and reddening also exists in 
the local universe (see Fig.~1 of Calzetti 2004), 
and was also found by Adelberger \& Steidel (2000) for $z\sim3$ 
LBGs, on a smaller range of reddening.

In Figures~\ref{fig:bzk_cor}c) and \ref{fig:bzk_cor}d), we plot 
the 
relation between color excess $E(B-V)$ and stellar mass of the
\sbzks\ in Daddi-F and Deep3a-F. The Spearman rank correlation
coefficient is $r_s= 0.53 $ for Daddi-F and $r_s= 0.63 $ for
Deep3a-F, indicating that the correlation between $E(B-V)$ and 
stellar mass is significant at the $>7\,\sigma$ level in both 
fields.  
In this case the estimate of the stellar mass depends only mildly 
on the assumed reddening, and therefore the correlation is likely 
to be intrinsic, with more massive galaxies being also more 
absorbed.

Given the previous two correlation, not surprisingly we also find
a correlation between SFR and stellar mass 
(Figure~\ref{fig:bzk_cor}e and \ref{fig:bzk_cor}f).
The Spearman rank correlation coefficient is $r_s= 0.30 $ for 
Daddi-F, and $r_s= 0.45 $ for Deep3a-F, indicating that the 
correlation between SFR and stellar mass is significant at the 
$>4\,\sigma$ level in both fields.  
The origin of the sharp edge in Fig.~\ref{fig:bzk_cor}e) and 
 Fig.~\ref{fig:bzk_cor}f) is caused by the color limit 
$BzK>-0.2$ for selecting \sbzks.  
To show this clearly, we plot galaxies with $-0.2<BzK<0.0$ as 
open squares in Fig.~\ref{fig:bzk_cor}f).  However, no or very 
few $z\sim2$ galaxies exist below the $BzK>-0.2$ line, and 
therefore the upper edge shown in the figure appears to be 
intrinsic, showing a limit on the maximum SFR that is likely to be 
present in a galaxy of a given mass.

At $z=0$ the vast majority of massive galaxies ($M_*\gsim
10^{11}M_\odot$) are passively evolving, ``red'' galaxies
(e.g., Baldry et al. 2004), while instead at $z\sim 2$, actively 
star-forming (\sbzks) and passive (\pegs) galaxies exist in similar
numbers, and Fig.~\ref{fig:bzk_cor} shows that the most massive
\sbzks\ tend also to be the most actively star forming. This can be
seen as yet another manifestation of the {\it downsizing} effect
(e.g., Cowie et al. 1996; Kodama et al. 2004; Treu et al. 2005), with
massive galaxies completing their star formation at an earlier epoch
compared to less massive galaxies, which instead have more prolonged 
star formation histories.

Because of the correlations discussed above, UV-selected samples
of $z\sim2$ galaxies (Steidel et al. 2004) will tend to preferentially
miss the most star-forming and most massive galaxies. 
Still, because of the large scattering in the correlations, some of
the latter galaxies will also be selected in the UV, as can be seen
in Fig.~\ref{fig:bzk_cor} and as emphasized in Shapley et al. (2004;
2005).

\subsection{Mass and SFR densities}

\begin{figure*}
\centering 
\includegraphics[angle=-90,width=0.9\textwidth]{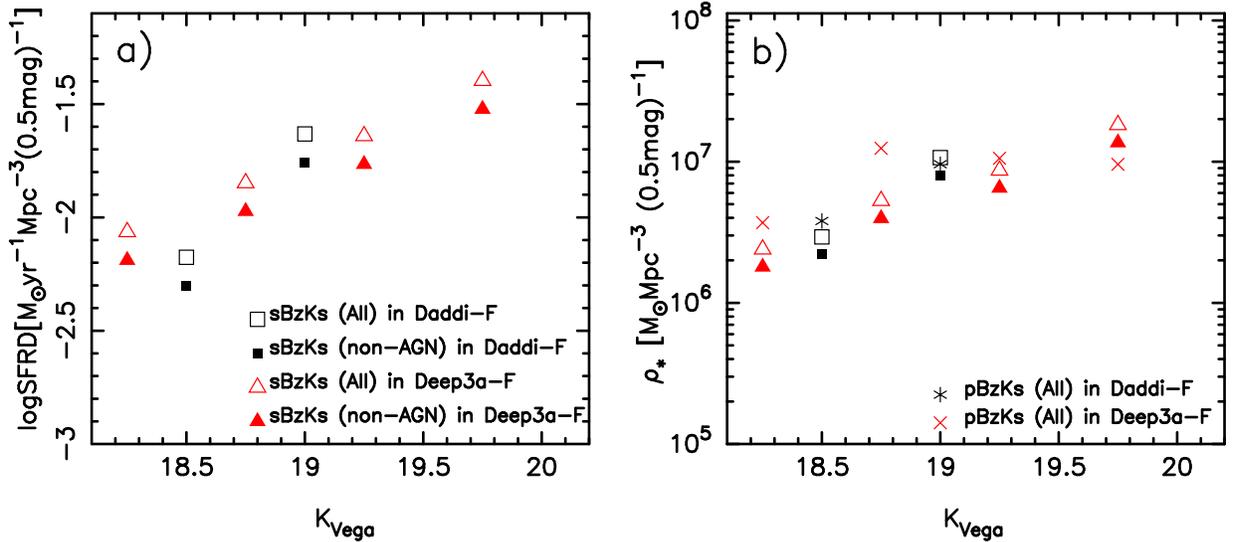}
\caption{
a) The differential contribution to the SFR density at $z\simeq 2$ 
from \sbzks\ as a function of their $K$-band magnitude. 
b) The differential contribution to the mass density at $z\simeq 2$ 
from \sbzks\ and \pegs\ as a function of their $K$-band magnitude. 
The open squares and triangles represent values were calculated 
from all \sbzks, the solid symbols represent values were corrected 
from the AGN contamination. The stars and crosses represent mass 
densities were calculated from \pegs. 
[{\it See the electronic edition of the Journal for the color version 
of this figure.}]}
\label{fig:sfrdmd}
\end{figure*}

In this subsection we derive the contribution of the \sbzks\ to 
the {\rm integrated} star formation rate density (SFRD) at 
$z\sim2$ and of \sbzks\ and \pegs\ to the stellar mass density 
at $z\sim2$. Some fraction of the \sbzks\ galaxies are known to 
be AGN-dominated galaxies (Daddi et al. 2004b; 2005b; Reddy et 
al. 2005).
To estimate the AGN contamination, we have used the 80 ks 
XMM-$Newton$ X-ray data that are available for Daddi-F (Brusa 
\etal 2005). 
A circular region of 11$'$ radius from the point of maximum 
exposure time include 70 \sbzks\, 18 of which are identified 
with X-ray sources using a 5${''}$ radius error circle (see 
Fig.~\ref{fig:bzk_cor}).  
This fraction is comparable to the one estimated in the CDFS field
(Daddi et al. 2004b) and in the GOODS-N field (Daddi et al. 2005b; 
Reddy et al. 2005), for which $>1$ Ms Chandra data are available. Based 
also on the latter result, we assume the AGN contamination is about 
25\%, and we adopt this fraction to statistically correct properties
derived from our \sbzks\ samples.

The left panel of Fig.~\ref{fig:sfrdmd} shows the differential
contribution to the SFR density at $z\simeq 2$ from \sbzks\ as a
function of their $K$-band magnitude.  Using the volume in the redshift
range $1.4 \lsim z \lsim 2.5$ (Daddi et al. 2004a; see also Reddy et al.
2005), an SFRD of $0.08$ $M_\sun$ yr$^{-1}$
Mpc$^{-3}$ is derived from the \sbzks\ ($\Kv < 20$) in Deep3a-F, and 
an
SFRD of $0.024$ $M_\sun$ yr$^{-1}$ Mpc$^{-3}$ is derived from the
\sbzks\ ($\Kv < 19.2$) in Daddi-F.  These estimates are reduced, 
respectively, to $0.06$ $M_\sun$ yr$^{-1}$ Mpc$^{-3}$ ($\Kv < 20$) and
0.018 $M_\sun$ yr$^{-1}$ Mpc$^{-3}$ ($\Kv < 19.2$) when subtracting an
estimated $25\%$ AGN contamination. Using the same method and the 24
\sbzks\ in the K20/GOODS-S sample, Daddi \etal (2004a) derived an SFRD
$~0.044\pm 0.008 $ $M_\sun$ yr$^{-1}$ Mpc$^{-3}$ for the volume in the
redshift range $1.4 \lsim z \lsim 2.5$; i.e., $\sim 25\%$ lower than
that derived here, possibly due to cosmic variance.  However, note that
there appears to be just a hint for the increasing trend in SFRD with
$K$ magnitude to flatten out at $\Kv\sim 20$, indicating that a
substantial contribution to the total SFRD is likely to come from $\Kv
> 20$ \sbzks, and therefore the values derived here should be regarded
as lower limits.  

Recently, Reddy et al. (2005) provided an estimate for the SFRD of 
$\Kv<20$ \sbzks\ of $\simlt0.02$ $M_\sun$ yr$^{-1}$ Mpc$^{-3}$, about 
one-third of the estimate that we have obtained here. 
Part of the difference is due to the absence of the reddest \sbzks\ 
from the Reddy et al. (2005) sample, as already noticed. 
However, most of the difference is likely due to the fact that the 
Reddy et al. (2005) SFR estimate is based primarily on the X-ray 
emission interpreted with the Ranalli et al. (2003) relation.
As shown by Daddi et al. (2005b), the X-ray emission interpreted in 
this way typically underestimates the SFR of \sbzks\ by factors of 2--3, 
with respect
to the radio-, mid-IR- and far-IR-based SFR estimates, 
all of which are also 
in reasonable agreement with the UV-corrected SFR estimate.

The right panel of Fig.~\ref{fig:sfrdmd} shows the differential
contribution to the stellar mass density $\rho_{\ast}$ at $z\simeq 2$ 
from \sbzks\ and \pegs\ as a function of their $K$-band magnitude.
The open squares and triangles represent values that were calculated 
from all \sbzks, the solid symbols represent values corrected 
for the AGN contamination. The stars and crosses represent the mass 
density contributed by \pegs. 

The stellar mass density in Deep3a-F, integrated to our $\Kv<20$
catalog limit, is 
log$\rho_*=7.7$ $M_\sun$ Mpc$^{-3}$, in excellent agreement with 
the value reported in Table 4 of Fontana \etal (2004), i.e.,
log$\rho_*=7.86$ $M_\sun$ Mpc$^{-3}$ for $1.5 \leq z <2.0$ 
galaxies, and log$\rho_*=7.65$ $M_\sun$ Mpc$^{-3}$ for 
$2.0 \leq z <2.5$ galaxies, but now from a much bigger sample.  
These estimates also agree with the log$\rho_*\sim 7.5$ $M_\sun$ 
Mpc$^{-3}$  estimate at $z\sim 2$ by Dickinson \etal\ (2003), 
although their selection is much deeper ($H_{\rm AB} <26.5$), 
although it extends 
over a much smaller field (HDF). So, while our sample is likely to 
miss the contribution of low-mass galaxies, the Dickinson \etal 
sample is likely to underestimate the contribution of high-mass 
galaxies due to the small field and cosmic variance. 
There is little evidence for flattening of 
log$\rho_*$ by $\Kv =20$. As already noted, the total stellar mass density 
at $z\sim2$ has to be significantly larger than that estimated here, i.e., 
only from the contribution of $\Kv <20$  $BzK$-selected galaxies.

There are  121 \pegs\ in Deep3a-F, and for $\sim 100$ of them we 
derive $M_* > 10^{11}$ $M_\sun$. Correspondingly, the number 
density of \pegs\ with $M_* > 10^{11}$ $M_\sun$ over the range 
$1.4 \lsim z \lsim 2.0$ is $(1.8\pm 0.2)\times10^{-4}$ Mpc$^{-3}$ 
(Poisson error only). 
This compares to $3.4\times10^{-4}$ Mpc$^{-3}$ over the same
redshift range as estimated by Daddi \etal (2005a) using six objects 
in the Hubble Ultra Deep Field (HUDF) with spectroscopic redshift.
While the Daddi et al. (2005a) HUDF sample is important to 
establish that most \pegs\ are indeed passively  evolving galaxies 
at $1.4<z<2.5$, their density measurements is fairly uncertain due 
to cosmic variance. 
Being derived from an area which is $\sim 30$ times larger than 
HUDF, the results presented here for the number density of massive, passively 
evolving galaxies in Deep3a-F in the quoted redshift range should 
be much less prone to cosmic variance. 
Hence, we estimate that, compared to the local value at 
$z\sim 0$ ($9\times10^{-4}$ Mpc$^{-3}$, Baldry \etal 2004), at 
$1.4<z<2$ there appears to be about 20\%$\pm$7\% 
of massive ($>10^{11}M_\odot$), passively evolving galaxies, with 
the error above accounting also for cosmic variance.

\section{Summary and Conclusions}\label{sec:summary}

This paper presents the results of a survey based on $BRIzJK$ 
photometry obtained by combining Subaru optical and ESO near-IR 
data over two separate fields (Deep3a-F and Daddi-F). 
Complete K-selected samples of galaxies were selected to $\Kv<20$
in the Deep3a-F over 320 arcmin$^2$, and to $\Kv\sim19$ in the Daddi-F
over a field roughly twice the  area.
Deep multicolor photometry in the $BRIz$ bands were obtained for the
objects in both fields.
Object catalogs constructed from these deep data contain more than
$10^4$ objects in the NIR bandpasses. Galaxy $K$-band number counts 
were derived and found to be in excellent agreement with previous survey 
results.

We have used color criteria to select candidate massive galaxies at high 
redshift, such as
$BzK$-selected star-forming (\sbzks) and passively evolving (\pegs)
galaxies at $1.4 \lsim z \lsim 2.5$, and EROs, 
and derived their number counts. The main results can be summarized 
as follows.

1. Down to the $K$-band limit of the survey the log of the number 
counts of \sbzks\ increases linearly with the $K$ magnitude, while 
that of \pegs\ flattens out by $\Kv \sim 19$. Over 
the Deep3a-F we select 387 \sbzks\ and 121 \pegs\ down to $\Kv=20$, 
roughly a factor of 10 more than over the 52 arcmin$^2$ fields of 
the K20 survey. This corresponds to a $\sim 30\%$ higher surface 
density, quite possibly the result of cosmic variance. Over Daddi-F 
we select 108  \sbzks\ and 48 \pegs\ down to $\Kv=19.2$.

2. The clustering properties (angular two-point correlation function) of
EROs and $BzK$-selected galaxies (both \sbzks\ and \pegs) are very
similar, and their clustering amplitudes are about a factor of 10
higher than those of generic galaxies in the same magnitude
range. The most strongly clustered populations at each redshift are
likely to be connected to each other in evolutionary terms, and therefore the
strong clustering of EROs and BzKs makes quite plausible an
evolutionary link between BzKs at $z\sim 2$ and EROs at $z\sim 1$,
with star formation in \sbzks\ subsiding by $z\sim 1$ thus producing
passively evolving EROs. While some \pegs\ may well experience
secondary, stochastic starbursts at lower redshift, the global
evolutionary trend of the galaxy population is dominated by star
formation being progressively quenched in massive galaxies, with the
quenching epoch of galaxies depending on environmental density, being
earlier in high-density regions.

3. Using approximate relations from Daddi et al. (2004a) and 
multicolor photometry, we estimated the color excess, SFR and 
stellar mass of \sbzks. 
These $K_{\rm Vega}<20$ galaxies have median reddening 
$E(B-V)\sim0.44$, average SFR$\sim190\ M_{\odot} yr^{-1}$, and 
typical stellar masses $\sim10^{11}M_\odot$.
Correlations between physical quantities are detected: the most massive
galaxies are those with the largest SFRs and optical reddening $E(B-V)$.
The high SFRs and masses of these galaxies add further support to 
the notion that these  $z\simeq2$ star-forming galaxies are among 
the precursors of $z\simeq1$ passive EROs and $z\simeq0$ early-type 
galaxies.

4. The contribution to the total star formation rate density at
$z\sim2$ was estimated for the $\Kv <20$ \sbzks\ in our fields. 
These vigorous starbursts produce an SFRD $\sim 0.06$ $M_\sun$ 
yr$^{-1}$ Mpc$^{-3}$, which is already comparable to the global SFRD at 
$z\sim 2$ as estimated from other surveys and simulations (e.g.
Springel \& Hernqwist 2003; Heavens et al. 2004).
However, a sizable additional contribution is expected from $\Kv >20$ 
\sbzks.

5. In a similar fashion, the stellar mass of \pegs\ was obtained, with
the result that the number density of $K_{\rm Vega}<20$ \pegs\ more
massive than $10^{11}M_\odot$ is about 20\%$\pm$7\% of that of similarly
massive, early-type galaxies at $z=0$, 
indicating that additional activity and subsequent quenching of 
star-formation in $\simgt10^{11} M_\odot$ star-forming galaxies must 
account for increasing the number of massive passive galaxies by a factor 
of about 5 from $z=1.7$. The number density of $\simgt10^{11} M_\odot$ 
$sBzK$s is similar to that of $pBzK$s. Given their strong star-formation 
activity, it seems that by $z\sim1$--1.4 the full population of local 
$\simgt10^{11} M_\odot$ passive galaxies could be eventually assembled as 
a result.

This result, advocated also in Daddi et al. (2005b), may appear in 
contradiction with the recent finding by Bell et al. (2004) of a 
factor of 2 decrease in the number density of early-type galaxies 
at $z\sim1$, with respect to the local value (see also Faber et al. 
2005). However, our analysis of the Bell et al (2004) results
shows that most of this evolution is to be ascribed to the 
progressive disappearence with increasing redshift of the fainter 
galaxies, while the population of the brightest, most massive 
galaxies being substantially stable. This would be, in fact, another 
manifestation of the {\it downsizing} effect.
A future publication will address this point in its full details and 
implications.

Mapping the metamorphosis of active
star-forming galaxies into passively evolving, early-type galaxies
from high to low redshifts, and as a function of galaxy mass and
environment is one of the primary goals of the main ongoing galaxy
surveys.
Using Subaru and VLT telescopes, optical and near-infrared spectra 
are being obtained, with targets from the present database having
been selected according to the same criteria adopted in this paper.  
Future 
papers in this series will present further scientific results from 
this {\it pilot} survey, along with a variety of data products.

\acknowledgments
We thank the anonymous referee for useful and constructive 
comments that resulted in a significant improvement of this paper.
The work is partly supported by a Grant-in-Aid for Scientific 
Research (16540223) by the Japanese Ministry of Education, 
Culture, Sports, Science and Technology and the Chinese National 
Science Foundation (10573014). 
X.K. gratefully acknowledges financial support from the JSPS. 
E.D. acknowledge support from NASA through the Spitzer Fellowship 
Program, under award 1268429.
L.F.O. thanks the Poincar\'{e} fellowship program at Observatoire de 
la C\^{o}te d'Azur and the Danish Natural Science Research Council 
for financial support.

\end{document}